\documentclass[journal, final]{IEEEtran}
\usepackage[utf8]{inputenc}
\usepackage[final]{graphicx}
\usepackage{subcaption}
\usepackage{mathtools}
\usepackage{amsmath,amstext,amsfonts,amssymb}
\usepackage{tikz}
\usepackage{pgfplots}
\usepackage{filecontents}
\usepackage{hyperref}
\usepackage[english]{babel}
\usepackage[nolist]{acronym} 

\newcommand{\bv}{{\bf b}}

\newcommand{\dv}{{\bf d}}

\newcommand{\hv}{{\bf h}}

\newcommand{\mv}{{\bf m}}
\newcommand{\nv}{{\bf n}}

\newcommand{\pv}{{\bf p}}

\newcommand{\rv}{{\bf r}}
\newcommand{\sv}{{\bf s}}

\newcommand{\uv}{{\bf u}}

\newcommand{\vv}{{\bf v}}
\newcommand{\xv}{{\bf x}}
\newcommand{\yv}{{\bf y}}

\newcommand{\zerov}{{\bf 0}}
\newcommand{\onev}{{\bf 1}}
\newcommand{\omegav}{{\boldsymbol \omega}}
\newcommand{\thetav}{{\boldsymbol \theta}}

\newcommand{\Id}{{\bf I}}

\newcommand{\Qm}{{\bf Q}}
\newcommand{\Rm}{{\bf R}}

\newcommand{\Wm}{{\bf W}}

\newcommand{\Xm}{{\bf X}}
\newcommand{\Ym}{{\bf Y}}

\newcommand{\Cc}{{\cal C}}

\newcommand{\Mc}{{\cal M}}
\newcommand{\Nc}{{\cal N}}

\newcommand{\Sc}{{\cal S}}

\newcommand{\CC}{\mathbb{C}}
\newcommand{\EE}{\mathbb{E}}

\newcommand{\RR}{\mathbb{R}}

\newcommand{\tp}{^{\sf T}}

\newcommand{\LB}{\left(}
\newcommand{\RB}{\right)}
\newcommand{\LSB}{\left[}
\newcommand{\RSB}{\right]}

\begin{acronym}
 \acro{ADC}{analog-to-digital converter}
 \acro{AGC}{automatic gain control}
 \acro{ASIC}{application-specific integrated circuit}
 \acro{AWGN}{additive white Gaussian noise}
 \acro{BC}{broadcast channel}
 \acro{BER}{bit error rate}
 \acro{BICM}{bit interleaved coded modulation}
 \acro{BLER}{block error rate}
 \acro{BPSK}{binary phase shift keying}
 \acro{CFO}{carrier frequency offset}
 \acro{CNN}{convolutional neural network}
 \acro{CSI}{channel state information}
 \acro{DL}{deep learning}
 \acro{DQPSK}{differential quadrature phase-shift keying}
 \acro{FPGA}{field programmable gate array}
 \acro{GNR}{GNU Radio}
 \acro{GPU}{graphic processing unit}
 \acro{ISI}{inter-symbol interference}
 \acro{MAC}{multiple access channel}
 \acro{MIMO}{multiple-input multiple-output}
 \acro{ML}{machine learning}
 \acro{MLP}{multilayer perceptron}
 \acro{MSE}{mean squared error}
 \acro{NN}{neural network}
 \acro{PA}{power amplifier}
 \acro{PLL}{phase-locked loop}
 \acro{ppm}{parts per million}
 \acro{PSK}{phase-shif keying}
 \acro{PFB}{polyphase filterbank}
 \acro{QAM}{quadrature amplitude modulation}
 \acro{QPSK}{quadrature phase shift keying}
 \acro{ReLU}{rectified linear unit}
 \acro{RNN}{recurrent neural network}
 \acro{RRC}{root-raised cosine}
 \acro{RTN}{radio transformer network}
 \acro{SDR}{software-defined radio}
 \acro{SFO}{sampling frequency offset}
 \acro{SGD}{stochastic gradient descent}
 \acro{SIMD}{single-instruction multiple-data}
 \acro{SNR}{signal-to-noise ratio}	
 \acro{TDL}{tapped delay line}
 \acro{t-SNE}{t-distributed stochastic neighbor embedding}
\end{acronym}

\begin{document}

\title{An Introduction to Deep Learning \\for the Physical Layer}

\author{Tim~O'Shea,~\IEEEmembership{Senior Member,~IEEE,} and Jakob~Hoydis,~\IEEEmembership{Member,~IEEE}%

\thanks{T.~O'Shea is with the Bradley Department of Electrical and Computer Engineering, Virginia Tech and DeepSig, Arlington, VA, US (oshea@vt.edu).}%

\thanks{J.~Hoydis is with Nokia Bell Labs, Route de Villejust, 91620 Nozay, France (jakob.hoydis@nokia-bell-labs.com).}}
\maketitle

\begin{abstract}
We present and discuss several novel applications of \ac{DL} for the physical layer. By interpreting a communications system as an autoencoder, we develop a fundamental new way to think about communications system design as an end-to-end reconstruction task that seeks to jointly optimize transmitter and receiver components in a single process. We show how this idea can be extended to networks of multiple transmitters and receivers and present the concept of \acp{RTN} as a means to incorporate expert domain knowledge in the \ac{ML} model. Lastly, we demonstrate the application of \acp{CNN} on raw IQ samples for modulation classification which achieves competitive accuracy with respect to traditional schemes relying on expert features. The paper is concluded with a discussion of open challenges and areas for future investigation.
\end{abstract}

\acresetall
\section{Introduction}
Communications is a field of rich expert knowledge about how to model channels of different types \cite{rappaport1996wireless, gagliardi1995optical}, compensate for various hardware imperfections \cite{meyr1998digital, schenk2008rf}, and design optimal signaling and detection schemes that ensure a reliable transfer of data \cite{proakis2007digital}. As such, it is a complex and mature engineering field with many distinct areas of investigation which have all seen diminishing returns with regards to performance improvements, in particular on the physical layer. Because of this, there is a high bar of performance over which any \ac{ML} or \ac{DL} based approach must pass in order to provide tangible new benefits.

In domains such as computer vision and natural language processing, \ac{DL} shines because it is difficult to characterize real world images or language with rigid mathematical models. For example, while it is an almost impossible task to write a robust algorithm for detection of handwritten digits or objects in images, it is almost trivial today to implement \ac{DL} algorithms that learn to accomplish this task beyond human levels of accuracy \cite{lecun89, he2015delving}. In communications, on the other hand, we can \emph{design} transmit signals that enable straightforward algorithms for symbol detection for a variety of
channel and system models (e.g., detection of a constellation symbol in \ac{AWGN}). Thus, as long as such models sufficiently capture real effects we do not expect \ac{DL} to yield significant improvements on the physical layer. 

Nevertheless, we believe that the \ac{DL} applications which we explore in this paper are a useful and insightful way of fundamentally rethinking the communications system design problem, and hold promise for performance improvements in complex communications scenarios that are difficult to describe with tractable mathematical models. Our main contributions are as follows:

\begin{itemize}
\item We demonstrate that it is possible to learn full transmitter and receiver implementations for a given channel model which are optimized for a chosen loss function (e.g., minimizing \ac{BLER}). Interestingly, such ``learned'' systems can be competitive with respect to the current state-of-the-art. The key idea here is to represent transmitter, channel, and receiver as one deep \ac{NN} that can be trained as an autoencoder. The beauty of this approach is that it can even be applied to channel models and loss functions for which the optimal solutions are unknown.

\item We extend this concept to an adversarial network of multiple transmitter-receiver pairs competing for capacity. This leads to the interference channel for which finding the best signaling scheme is a long-standing research problem. We demonstrate that such a setup can also be represented as an \ac{NN} with multiple inputs and outputs, and that all transmitter and receiver implementations can be jointly optimized with respect to a common or individual performance metric(s).

\item We introduce \acp{RTN} as a way to integrate expert knowledge into the \ac{DL} model. \acp{RTN} allow, for example, to carry out predefined correction algorithms (``transformers'') at the receiver (e.g., multiplication by a complex-valued number, convolution with a vector) which may be fed with parameters learned by another \ac{NN}. This \ac{NN} can be integrated into the end-to-end training process of a task performed on the transformed signal (e.g., symbol detection).

\item We study the use of \acp{NN} on complex-valued IQ samples for the problem of modulation classification and show that \acp{CNN}, which are the cornerstone of most \ac{DL} systems for computer vision, can outperform traditional classification techniques based on expert features.  This result mirrors a relentless trend in \ac{DL} for various domains, where learned features ultimately outperform and displace long-used expert features, such as the scale-invariant feature transform (SIFT) \cite{lowe1999object} and Bag-of-words \cite{harris1954distributional}.
\end{itemize}
 
The ideas presented in this paper provide a multitude of interesting avenues for future research that will be discussed in detail. We hope that these will stimulate wide interest within the research community.

The rest of this article is structured as follows:
Section~\ref{ssec:can_dl} discusses potential benefits of \ac{DL} for the physical layer. Section~\ref{ssec:background} presents related work. Background of deep learning is presented in Section~\ref{sec:dl}. In Section~\ref{sec:ml-applications}, several \ac{DL} applications for communications are presented. Section~\ref{sec:discussion} contains an overview and discussion of open problems and key areas of future investigation. Section~\ref{sec:conclusions} concludes the article.

\subsection{Potential of \ac{DL} for the physical layer}\label{ssec:can_dl}
Apart from the intellectual beauty of a fully ``learned'' communications system, there are some reasons why \ac{DL} could provide gains over existing physical layer algorithms.

First, most signal processing algorithms in communications have solid foundations in statistics and information theory and are often provably optimal for tractable mathematically models. These are generally linear, stationary, and have Gaussian statistics. A practical system, however, has many imperfections and non-linearities \cite{schenk2008rf} (e.g., non-linear \acp{PA}, finite resolution quantization) that can only be approximately captured by such models. For this reason, a \ac{DL}-based communications system (or processing block) that does not require a mathematically tractable model and that can be optimized for a specific hardware configuration and channel might be able to better optimize for such imperfections. 

Second, one of the guiding principles of communications systems design is to split the signal processing into a chain of multiple independent blocks; each executing a well defined and isolated function (e.g., source/channel coding, modulation, channel estimation, equalization). Although this approach has led to the efficient, versatile, and controllable systems we have today, it is not clear that individually optimized processing blocks achieve the best possible end-to-end performance. For example, the separation of source and channel coding for many practical channels and short block lengths (see \cite{goldsmith1995joint} and references therein) as well as separate coding and modulation \cite{zehavi19928} are known to be sub-optimal. Attempts to jointly optimize each of these components, e.g., based on factor graphs \cite{wymeersch2007iterative}, provide gains but lead to unwieldy and computationally complex systems.  A learned end-to-end communications system on the other hand is unlikely to have such a rigid modular structure as it is optimized for end-to-end performance. 

Third, it has been shown that \acp{NN} are universal function approximators \cite{hornik1989multilayer} and recent work has shown a remarkable capacity for algorithmic learning with recurrent \acp{NN} \cite{reed2015neural} that are known to be Turing-complete \cite{siegelmann1992computational}.  Since the execution of \acp{NN} can be highly parallelized on concurrent architectures and easily implemented with low-precision data types \cite{vanhoucke2011improving}, there is evidence that ``learned'' algorithms taking this form could be executed faster and at lower energy cost than their manually ``programmed'' counterparts. 

Fourth, massively parallel processing architectures with distributed memory architectures, such as \acp{GPU} but also increasingly specialized chips for \ac{NN} inference (e.g., \cite{chen2017eyeriss}), have shown to be very energy efficient and capable of impressive computational throughput when fully utilized by concurrent algorithms \cite{raina2009large}. The performance of such architectures, however, has been largely limited by the ability of algorithms and higher level programming languages to make efficient use of them. The inherently concurrent nature of computation and memory access across wide and deep \acp{NN} has demonstrated a surprising ability to readily achieve high resource utilization on these architectures with minimal application specific tuning or optimization required.

\subsection{Historical context and related work}\label{ssec:background}

Applications of \ac{ML} in communications have a long history covering a wide range of applications. These comprise channel modeling and prediction, localization, equalization, decoding, quantization, compression, demodulation, modulation recognition, and spectrum sensing to name a few \cite{ibnkahla2000applications, bkassiny2013survey} (and references therein). However, to the best of our knowledge and due to the reasons mentioned above, few of these applications have been commonly adopted or led to a wide commercial success. It is also interesting that essentially all of these applications focus on individual receiver processing tasks alone, while the consideration of the transmitter or a full end-to-end system is entirely missing in the literature. 

The advent of open-source \ac{DL} libraries (see Section~\ref{sec:ml-lib}) and readily available specialized hardware along with the astonishing progress of \ac{DL} in computer vision have stimulated renewed interest in the application of \ac{DL} for communications and networking \cite{qadir2015ieee}. There are currently essentially two different main approaches of applying \ac{DL} to the physical layer. The goal is to either improve/augment parts of existing algorithms with \ac{DL}, or to completely replace them.

Among the papers falling into the first category are \cite{nachmani2016learning, nachmani2017rnn} and \cite{samuel2017deep} that consider improved belief propagation channel decoding and \ac{MIMO} detection, respectively. These works are inspired by the idea of \emph{deep unfolding} \cite{hershey2014deep} of existing iterative algorithms by essentially interpreting each iteration as a set of \ac{NN} layers. In a similar manner, \cite{borgerding2016onsager} aims at improving the solution of sparse linear inverse problems with \ac{DL}.

In the second category, papers include \cite{jeon2016blind}, dealing with blind detection for \ac{MIMO} systems with low-resolution quantization, and \cite{farsad2017detection}, in which detection for molecular communications for which no mathematical channel model exists is studied. The idea of learning to solve complex optimization tasks for wireless resource allocation, such as power control, is investigated in \cite{sun2017learning}. Some of us have also demonstrated initial results in the area of learned end-to-end communications systems \cite{o2016learning} as well as considered the problems of modulation recognition \cite{o2016convolutional}, signal compression \cite{o2016unsupervised}, and channel decoding \cite{gruber2017, cammerer2017scaling} with state-of-the art \ac{DL} tools. 

\textbf{Notations:}
We use boldface upper- and lower-case letters to denote matrices and column vectors, respectively. For a vector $\xv$, $x_i$ denotes its $i$th element, $\|\xv\|$ its Euclidean norm, $\xv\tp$ its transpose, and $\xv \odot \yv$  the element-wise product with $\yv$. For a matrix $\Xm$, $X_{ij}$ or $[\Xm]_{ij}$ denotes the $(i,j)$-element. $\RR$ and $\CC$ denote the sets of real and complex numbers, respectively. $\Nc(\mv,\Rm)$ and $\Cc\Nc(\mv,\Rm)$ are the multivariate  Gaussian and complex Gaussian distributions with mean vector $\mv$ and covariance matrix $\Rm$, respectively. $\operatorname{Bern}(\alpha)$ is the Bernoulli distribution with success probability $\alpha$ and $\nabla$ is the gradient operator.

\section{Deep learning basics}\label{sec:dl}
A feedforward \ac{NN} (or \ac{MLP}) with $L$ layers describes a mapping $f(\rv_0; \thetav): \RR^{N_{0}} \mapsto \RR^{N_{L}}$ of an input vector ${\rv_0\in\RR^{N_0}}$ to an output vector ${\rv_L\in\RR^{N_L}}$ through $L$ iterative processing steps:
\begin{align}
\rv_\ell = f_\ell(\rv_{\ell-1}; \theta_\ell), \qquad \ell=1,\dots,L
\end{align}
where $f_\ell(\rv_{\ell-1};\theta_\ell): \RR^{N_{\ell-1}} \mapsto \RR^{N_{\ell}}$ is the mapping carried out by the $\ell$th layer. This mapping depends not only on the output vector $\rv_{\ell-1}$ from the previous layer but also on a set of parameters $\theta_\ell$. Moreover, the mapping can be stochastic, i.e., $f_\ell$ can be a function of some random variables. We use ${\thetav=\{\theta_1,\dots,\theta_L\}}$ to denote the set of all parameters of the network. The $\ell$th layer is called \emph{dense} or \emph{fully-connected} if $f_\ell(\rv_{\ell-1}; \theta_\ell)$ has the form
\begin{align}\label{eq:dense-layer}
f_\ell(\rv_{\ell-1}; \theta_\ell) = \sigma\left( \Wm_\ell \rv_{\ell-1} + \bv_\ell \right)
\end{align}
where $\Wm_\ell\in\RR^{N_\ell\times N_{\ell-1}}$, $\bv_\ell\in\RR^{N_\ell}$,
and $\sigma(\cdot)$ is an \emph{activation} function which we will be defined shortly. The set of parameters for this layer is  $\theta_\ell=\{\Wm_\ell, \bv_\ell\}$. Table~\ref{tab:layer-types} lists several other layer types together with their mapping functions and parameters which are used in this manuscript. All layers with stochastic mappings generate a new random mapping each time they are called. For example, the noise layer simply adds a Gaussian noise vector with zero mean and covariance matrix $\beta\Id_{N_{\ell-1}}$ to the input. Thus, it generates a different output for the same input each time it is called. The activation function $\sigma(\cdot)$ in \eqref{eq:dense-layer} introduces a non-linearity which is important for the so-called \emph{expressive power} of the \ac{NN}. Without this non-linearity there would be not much of an advantage of stacking multiple layers on top of each other. Generally, the activation function is applied individually to each element of its input vector, i.e.,  $[\sigma(\uv)]_i = \sigma(u_i)$. Some commonly used activation functions are listed in Table~\ref{tab:activation-functions}.\footnote{The linear activation function is typically used at the output layer in the context of regression tasks, i.e., estimation of a real-valued vector.} \acp{NN} are generally trained using labeled training data, i.e., a set of input-output vector pairs $(\rv_{0,i}, \rv^\star_{L,i})$, $i=1,\dots,S$, where $\rv^\star_{L,i}$ is the desired output of the neural network when $\rv_{0,i}$ is used as input. The goal of the training process is to minimize the loss 
\begin{align}
L(\thetav) = \frac1S\sum_{i=1}^S l(\rv^\star_{L,i},\rv_{L,i})
\end{align}
with respect to the parameters in $\thetav$, where ${l(\uv,\vv): \RR^{N_L}\times \RR^{N_L} \mapsto \RR}$ is the loss function and $\rv_{L,i}$ is the output of the \ac{NN} when $\rv_{0,i}$ is used as input. Several relevant loss functions are provided in Table~\ref{tab:loss-functions}. Different norms (e.g., $L1$, $L2$) of parameters or activations can be added to the loss function to favor solutions with small or sparse values (a form of regularization). The most popular algorithm to find good sets of parameters $\thetav$ is \ac{SGD} which starts with some random initial values of $\thetav = \thetav_0$ and then updates $\thetav$ iteratively as
\begin{align}\label{eq:sgd}
\thetav_{t+1} = \thetav_t - \eta  \nabla\tilde{L}(\thetav_t)
\end{align}
where $\eta>0$ is the learning rate and $\tilde{L}(\thetav)$ is an approximation of the loss function which is computed for a random \emph{mini-batch} of training examples $\Sc_t \subset \{1,2,\dots,S\}$ of size $S_t$ at each iteration, i.e.,
\begin{align}\label{eq:def_loss}
\tilde{L}(\thetav) = \frac1{S_t}\sum_{i\in \Sc_t} l(\rv^\star_{L,i},\rv_{L,i}).
\end{align}
By choosing $S_t$ small compared to $S$, the gradient computation complexity is significantly reduced while still reducing weight update variance. Note that there are many variants of the \ac{SGD} algorithm which dynamically adapt the learning rate to improve convergence \cite[Ch.~8.5]{Goodfellow-et-al-2016-Book}. The gradient in \eqref{eq:sgd} can be very efficiently computed through the backpropagation algorithm \cite[Ch.~6.5]{Goodfellow-et-al-2016-Book}. Definition and training of \acp{NN} of almost arbitrary shape can be easily done with one of the many existing \ac{DL} libraries presented in Section~\ref{sec:ml-lib}.

\begin{table}
\renewcommand{\arraystretch}{1.2} 
\centering
\caption{List of layer types}
\label{tab:layer-types}
\begin{tabular}{c|c|c}
Name    & $f_\ell(\rv_{\ell-1}; \theta_\ell)$              & $\theta_\ell$        \\\hline
Dense    & $\sigma\left( \Wm_\ell \rv_{\ell-1} + \bv_\ell \right)$                    & $\Wm_\ell, \bv_\ell$ \\
Noise & $\rv_{\ell-1} + \nv$, $\nv\sim\Nc(\zerov,\beta\Id_{N_{\ell-1}})$           & none     \\
Dropout \cite{srivastava2014dropout} & $\dv \odot \rv_{\ell-1}$, $d_i\sim \operatorname{Bern}(\alpha)$ & none  \\
Normalization & e.g., $\frac{\sqrt{N_{\ell-1}}\rv_{\ell-1}}{\lVert \rv_{\ell-1} \rVert_2}$ &    none
\end{tabular}
\end{table}

\begin{table}
\renewcommand{\arraystretch}{1.2} 
\centering
\caption{List of activation functions}
\label{tab:activation-functions}
\begin{tabular}{c|c|c}
Name    & $[\sigma(\bf u)]_i$              & Range        \\\hline
linear & $u_i$ & $(-\infty,\infty)$ \\
ReLU \cite{nair2010rectified}    & $\max(0,u_i)$                    & $[0,\infty)$ \\
tanh    & $\tanh(u_i)$                      & $(-1,1)$\\
sigmoid & $\frac{1}{1+e^{-u_i}}$           & $(0,1)$     \\
softmax & $\frac{e^{u_i}}{\sum_j e^{u_j}}$ & $(0,1)$     
\end{tabular}
\end{table}

\begin{table}[t!]
\renewcommand{\arraystretch}{1.2} 
\centering
\caption{List of loss functions}
\label{tab:loss-functions}
\begin{tabular}{c|c}
Name    & $l(\uv,\vv)$    \\\hline
MSE    & $\lVert\uv - \vv\rVert_2^2$ \\
Categorical cross-entropy & $-\sum_{j} u_j\log(v_j)$     
\end{tabular}
\end{table}

\subsection{Convolutional layers}
Convolutional neural network (CNN) layers were introduced in \cite{lecun89} to provide an efficient learning method for 2D images.  By tying adjacent shifts of the same weights together in a way similar to that of a filter sliding across an input vector, convolutional layers are able to force the learning of features with an invariance to shifts in the input vector. In doing so, they also greatly reduce the model complexity (as measured by the number of free parameters in the layer's weight matrix) required to represent equivalent shift-invariant features using fully connected layers, reducing \ac{SGD} optimization complexity and improving generalization on appropriate datasets.

In general, a convolutional layer consists of a set of $F$ filter weights $\Qm^f \in \RR^{a \times b}$, $f=1,\dots,F$ ($F$ is called the \emph{depth}), which generate each a so-called \emph{feature map} $\Ym^f \in \RR^{n' \times m'} $ from an input matrix $\Xm \in \RR^{n \times m}$ \footnote{In image processing, $\Xm$ is commonly a three-dimensional tensor with the third dimension corresponding to color channels. The filters weights are also three-dimensional and work on each input channels simultaneously.} according to the following convolution:
\begin{equation}
Y^f_{i,j} = \sum_{k=0}^{a-1} \sum_{\ell=0}^{b-1} Q^f_{a-k,b-\ell} X_{1+s(i-1)-k,1+s(j-1)-\ell}
\end{equation}
where $s\geq 1$ is an integer parameter called \emph{stride}, $n'=1+\lfloor \frac{n+a-2}{s}\rfloor$ and $m'=1+\lfloor \frac{m+b-2}{s}\rfloor$,  and it is assumed that $\Xm$ is padded with zeros, i.e., $X_{i,j}=0$ for all ${i\notin[1,n]}$ and ${j\notin[1,m]}$. The output dimensions can be reduced by either increasing the stride $s$ or by adding a \emph{pooling} layer. The pooling layer partitions $\Ym$ into $p\times p$ regions for each of which it computes a single output value, e.g., maximum or average value, or $L2$-norm. 

For example, taking a vectorized grayscale image input consisting of $28\times 28$ pixels and connecting it to a dense layer with the same number of activations, results in a single weight matrix with $784 \times 784 = 614,656$ free parameters.   On the other hand, if we use a convolutional feature map containing six filters each sized $5\times 5$ pixels, we obtain a much reduced number of free parameters of $6\cdot 5 \cdot 5 = 150$.  For the right kind of dataset, this technique can be extremely effective. We will see an application of convolutional layers in Section~\ref{ssec:classification}. For more details on \acp{CNN}, we refer to \cite[Ch.~9]{Goodfellow-et-al-2016-Book}.

\subsection{Machine learning libraries}\label{sec:ml-lib}
In recent times, numerous tools and algorithms have emerged that make it easy to build and train large \acp{NN}. Tools to deploy such training routines from high level language to massively parallel \ac{GPU} architectures have been key enablers. Among these are Caffe \cite{jia2014caffe}, MXNet \cite{chen2015mxnet}, TensorFlow \cite{tensorflow2015-whitepaper}, Theano \cite{theano}, and Torch \cite{torch} (just to name a few), which allow for high level algorithm definition in various programming languages or configuration files, automatic differentiation of training loss functions through arbitrarily large networks, and compilation of the network's forwards and backwards passes into hardware optimized concurrent dense matrix algebra kernels. Keras \cite{chollet2015} provides an additional layer of \ac{NN} primitives with Theano and TensorFlow as its back-end. It has a highly customizable interface to quickly experiment with and deploy deep \acp{NN}, and has become our primary tool used to generate the numerical results for this manuscript \cite{radioML}.

\subsection{Network dimensions and training}
\label{ssec:training}
The term ``deep'' has become common in recent \ac{NN} literature, referring to the number of sequential layers within a network (but also more generally to the methods commonly used to train such networks). Depth relates directly to the number of iterative operations performed on input data through sequential layers' transfer functions. While deep networks allow for numerous iterative transforms on the data, a minimum latency network would likely be as shallow as possible. ``Width'' is used to describe the number of output activations per layer, or for all layers on average, and relates directly to the memory required by each layer.

Best practice training methods have varied over the years, from direct solution techniques over gradient descent to genetic algorithms, each having been favored or considered at one time (see \cite[Ch.~1.2]{Goodfellow-et-al-2016-Book} for a short history of \ac{DL}). Layer-by-layer pre-training \cite{hinton2006fast} was also a recently popular method for scaling training to larger networks where backpropagation once struggled. However, most systems today are able to train networks which are both wide and deep directly using backpropagation and \ac{SGD} methods with adaptive learning rates (e.g., Adam \cite{kingma2014adam}), regularization methods to prevent overfitting (e.g., Dropout \cite{srivastava2014dropout}), and activations functions which reduce gradient issues (e.g., ReLU \cite{nair2010rectified}).

\section{Examples of machine learning applications for the physical layer}\label{sec:ml-applications}
In this section, we will show how to represent an end-to-end communications system as an autoencoder and train it via \ac{SGD}. This idea is then extended to multiple transmitters and receivers and we study as an example the two-user interference channel. We will then introduce the concept of \acp{RTN} to improve performance on fading channels, and demonstrate the application of \acp{CNN} to raw radio frequency time-series data for the task of modulation classification.

\subsection{Autoencoders for end-to-end communications systems}\label{ssec:ae}
\begin{figure}[!h]
\centering
\includegraphics[width=0.45\textwidth]{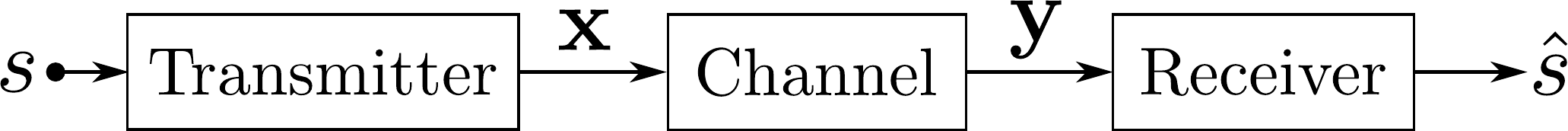}
\caption{A simple communications system consisting of a transmitter and a receiver connected through a channel\label{fig:communications_system}}
\end{figure}

\begin{figure*}
\centering
\includegraphics[width=0.65\textwidth]{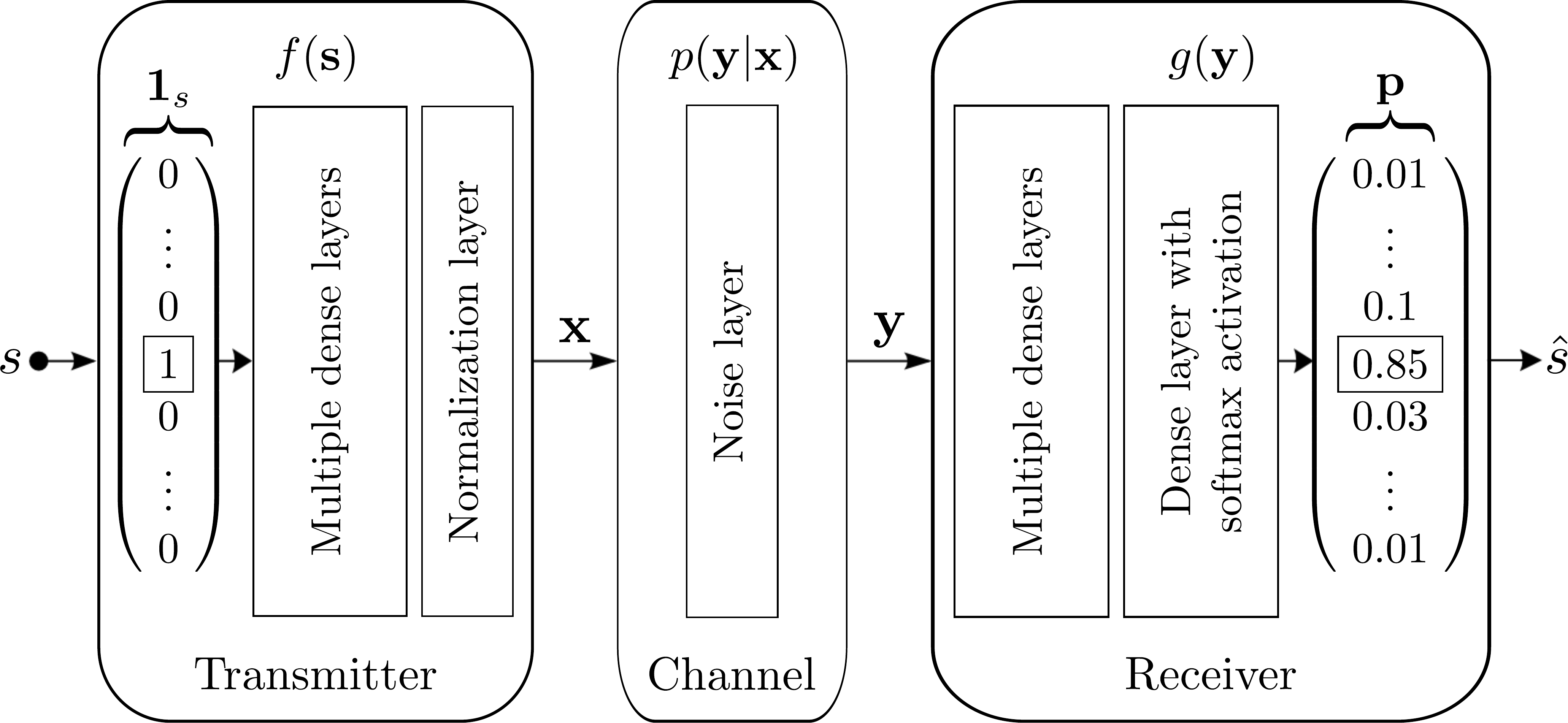}
\caption{A communications system over an AWGN channel represented as an autoencoder. The input $s$ is encoded as a one-hot vector, the output is a probability distribution over all possible messages from which the most likely is picked as output $\hat{s}$. \label{fig:ae_awgn}}
\end{figure*} 

In its simplest form, a communications system consists of a transmitter, a channel, and a receiver, as shown in Fig.~\ref{fig:communications_system}. The transmitter wants to communicate one out of $M$ possible messages $s\in \Mc=\{1,2,...,M\}$ to the receiver making $n$ discrete uses of the channel. To this end, it applies the transformation $f: \Mc \mapsto \RR^{n}$ to the message $s$ to generate the transmitted signal $\xv=f(s)\in\RR^{n}$.\footnote{We focus here on real-valued signals only. Extensions to complex-valued signals are discussed in Section~\ref{sec:discussion}. Alternatively, one can consider a mapping to $\RR^{2n}$, which can be interpreted as a concatenation of the real and imaginary parts of $\xv$. This approach is adopted in Sections~\ref{ssec:interference_channel}, \ref{ssec:rtn}, and \ref{ssec:classification}.} Generally, the hardware of the transmitter imposes certain constraints on $\xv$, e.g., an energy constraint $\lVert \xv \rVert_2^2 \le n$, an amplitude constraint $| x_i | \le 1\, \forall i$, or an average power constraint $\EE\LSB |x_i|^2 \RSB\le 1\,\forall i $. The communication rate of this communications system is $R=k/n$\,[bit/channel~use], where $k=\log_2(M)$. In the sequel, the notation ($n$,$k$) means that a communications system sends one out of $M=2^k$ messages (i.e., $k$ bits) through $n$ channel uses. The channel is described by the conditional probability density function $p(\yv|\xv)$, where $\yv\in\RR^n$ denotes the received signal. Upon reception of $\yv$, the receiver applies the transformation $g:\RR^n \mapsto \Mc$ to produce the estimate $\hat{s}$ of the transmitted message $s$.
  
From a \ac{DL} point of view, this simple communications system can be seen as a particular type of \emph{autoencoder} \cite[Ch.~14]{Goodfellow-et-al-2016-Book}. Typically, the goal of an autoencoder is to find a low-dimensional representation of its input at some intermediate layer which allows reconstruction at the output with minimal error. In this way, the autoencoder learns to non-linearly compress and reconstruct the input. In our case, the purpose of the autoencoder is different. It seeks to learn representations $\xv$ of the messages $s$ that are robust with respect to the channel impairments mapping $\xv$ to $\yv$ (i.e., noise, fading, distortion, etc.), so that the transmitted message can be recovered with small probability of error. In other words, while most autoencoders remove redundancy from input data for compression, this autoencoder (the ``channel autoencoder'') often adds redundancy, learning an intermediate representation robust to channel perturbations.

An example of such an autoencoder is shown in Fig.~\ref{fig:ae_awgn}. Here, the transmitter consists of a feedforward \ac{NN} with multiple dense layers followed by a normalization layer that ensures that physical constraints on $\xv$ are met. Note that the input $s$ to the transmitter is encoded as a \emph{one-hot vector} $\onev_s\in\RR^{M}$, i.e., an $M$-dimensional vector, the $s$th element of which is equal to one and zero otherwise. The channel is represented by an additive noise layer with a fixed variance $\beta=(2RE_b/N_0)^{-1}$, where $E_b/N_0$ denotes the energy per bit ($E_b$) to noise power spectral density ($N_0$) ratio. The receiver is also implemented as a feedforward \ac{NN}. Its last layer uses a softmax activation whose output $\pv\in(0,1)^{M}$ is a probability vector over all possible messages. The decoded message $\hat{s}$ corresponds then to the index of the element of $\pv$ with the highest probability. The autoencoder can then be trained end-to-end using \ac{SGD} on the set of all possible messages $s\in\Mc$ using the well suited categorical cross-entropy loss function between $\onev_s$ and $\pv$.\footnote{A more memory-efficient approach to implement this architecture is by replacing the one-hot encoded input and the first dense layer by an \emph{embedding} that turns message indices into vectors. The loss function can then be replaced by the \emph{sparse categorical cross-entropy} that accepts message indices rather than one-hot vectors as labels. This was done in our experiments \cite{radioML}.}

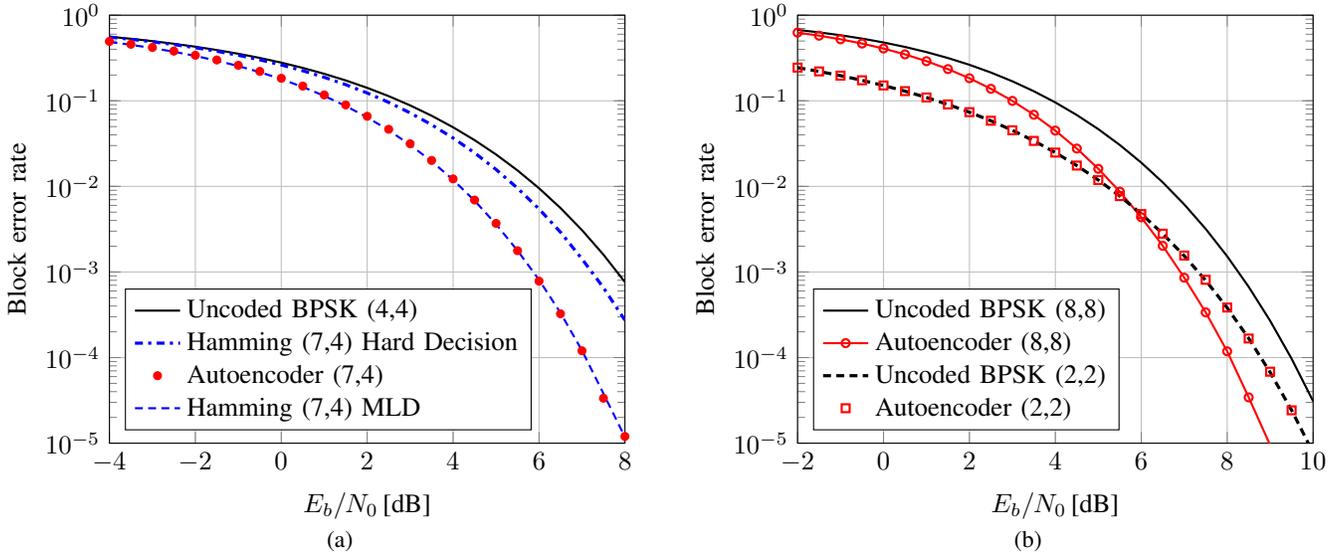
\begin{figure*}[t!]
\begin{subfigure}{0.5\textwidth}
 \begin{tikzpicture}
  \begin{semilogyaxis}[
    xmin=-4,
    xmax=8,
    ymin=0.00001,
    ymax=1,
    grid=major,
    xlabel={$E_b/N_0$\,[dB]},
    ylabel={Block error rate},
    legend pos=south west,
    legend cell align={left},
    ]
   \addplot[black, thick] 
    table [x=xa, y=ya, col sep=comma]{autoencoder_vs_hamming.csv};
   \addplot[blue, very thick, dashdotted] 
    table [x=xb, y=yb, col sep=comma]{autoencoder_vs_hamming.csv};
   \addplot[red, only marks, mark size={1.5}] 
    table [x=xc, y=yc, col sep=comma]{autoencoder_vs_hamming.csv};
   \addplot[blue, densely dashed, thick] 
    table [x=xd, y=yd, col sep=comma]{autoencoder_vs_hamming.csv};
   \legend{{Uncoded BPSK (4,4)},{Hamming (7,4) Hard Decision}, {Autoencoder (7,4)}, {Hamming (7,4) MLD}}
  \end{semilogyaxis}
 \end{tikzpicture}\vskip-2mm
 \caption{}\label{fig:nn_vs_hamming}
\end{subfigure}
\begin{subfigure}{0.5\textwidth}
 \begin{tikzpicture}
  \begin{semilogyaxis}[
    xmin=-2,
    xmax=10,
    ymin=0.00001,
    ymax=1,
    grid=major,
    xlabel={$E_b/N_0$\,[dB]},
    ylabel={Block error rate},
    legend pos=south west,
    legend cell align={left},
    ]
   \addplot[black, thick] 
    table [x=xa, y=ya, col sep=comma]{autoencoder_vs_bpsk_8_8.csv};
   \addplot[red, mark=o, mark size={1.5}, thick] 
    table [x=xb, y=yb, col sep=comma]{autoencoder_vs_bpsk_8_8.csv};   
   \addplot[black, densely dashed, very thick] 
    table [x=xa, y=ya, col sep=comma]{autoencoder_vs_bpsk_2_2.csv};
   \addplot[red, only marks, mark size={1.5}, mark=square, thick] 
    table [x=xb, y=yb, col sep=comma]{autoencoder_vs_bpsk_2_2.csv};      
   \legend{{Uncoded BPSK (8,8)}, {Autoencoder (8,8)}, {Uncoded BPSK (2,2)}, {Autoencoder (2,2)}}
  \end{semilogyaxis}
 \end{tikzpicture}\vskip-2mm
 \caption{}\label{fig:nn_vs_bpsk}
 \end{subfigure}
 \caption{BLER versus $E_b/N_0$ for the autoencoder and several baseline communication schemes}
\end{figure*}

Fig.~\ref{fig:nn_vs_hamming} compares the block error rate (\ac{BLER}), i.e., ${\Pr(\hat{s}\neq s)}$, of a communications system employing binary phase-shift keying (BPSK) modulation and a Hamming~(7,4) code with either binary hard-decision decoding or maximum likelihood  decoding (MLD) against the \ac{BLER} achieved by the trained autoencoder~(7,4) (with fixed energy constraint $\lVert \xv \rVert_2^2 = n$). Both systems operate at rate $R=4/7$. For comparison, we also provide the \ac{BLER} of uncoded BPSK~(4,4). This result shows that the autoencoder has learned without any prior knowledge an encoder and decoder function that together achieve the same performance as the Hamming~(7,4) code with MLD. The layout of the autoencoder is provided in Table~\ref{tab:ae-layout}. 
Although a single layer can represent the same mapping from message index to corresponding transmit vector, our experiments have shown that \ac{SGD} converges to a better global solution using two transmit layers instead of one. This increased dimension parameter search space may actually help to reduce likelihood of convergence to sub-optimal minima by making such solutions more likely to emerge as saddle points during optimization \cite{dauphin2014identifying}.
Training was done at a fixed value of $E_b/N_0=7\,$dB (cf. Section~\ref{ssec:data-loss-snr}) using Adam \cite{kingma2014adam} with learning rate $0.001$. We have observed that increasing the batch size during training helps to improve accuracy. For all other implementation details, we refer to the source code \cite{radioML}.

Fig.~\ref{fig:nn_vs_bpsk} shows a similar comparison but for an (8,8) and (2,2) communications system, i.e., $R=1$. Surprisingly, while the autoencoder achieves the same \ac{BLER} as uncoded BPSK for (2,2), it outperforms the latter for (8,8) over the full range of $E_b/N_0$. This implies that it has learned some joint coding and modulation scheme, such that a coding gain is achieved. For a truly fair comparison, this result should be compared to a higher-order modulation scheme using a channel code (or the optimal sphere packing in eight dimensions). A detailed performance comparison for various channel types and parameters $(n,k)$ with different baselines is out of the scope of this paper and left to future investigations.

\begin{figure}
\centering
\begin{tabular}{c}
\includegraphics[width=0.22\textwidth]{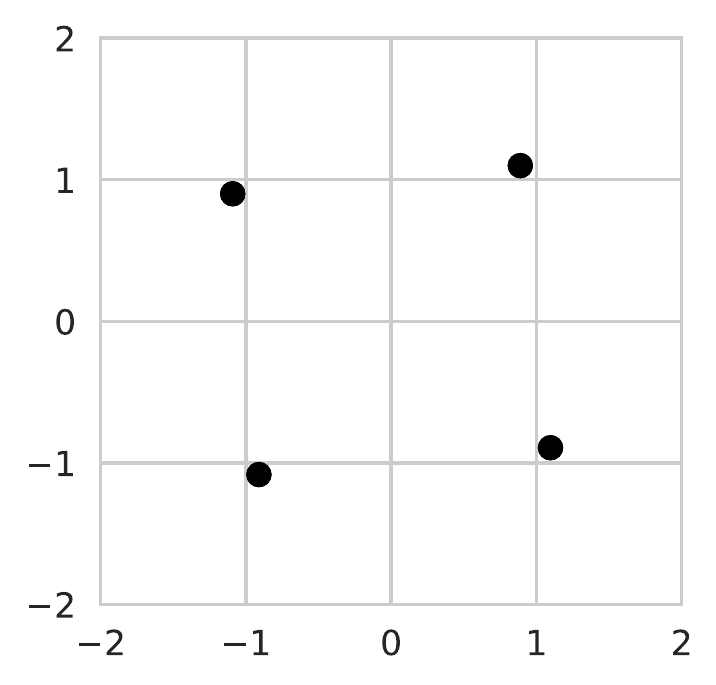}        
 \includegraphics[width=0.22\textwidth]{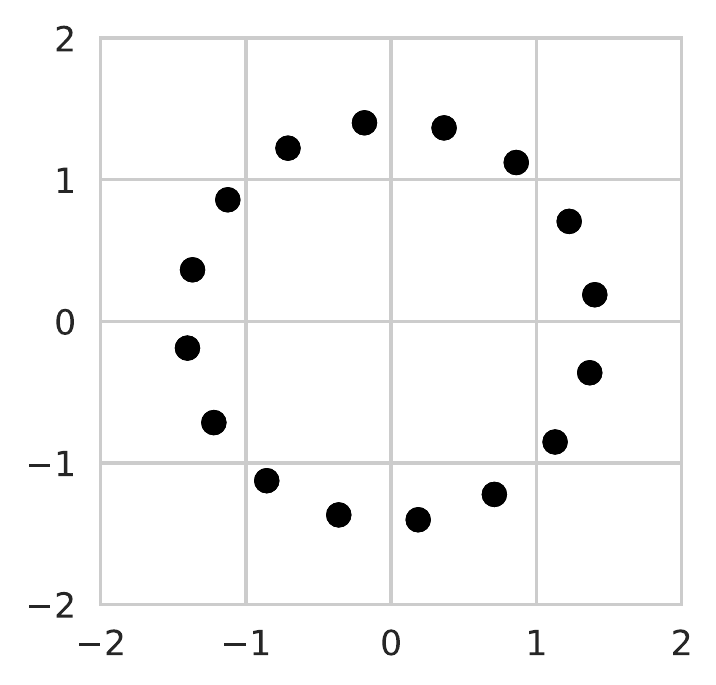}        
\\[-5pt]   
\footnotesize{\hskip4.3mm(a)} \hskip36.8mm  \footnotesize{(b)} 
 \\
\includegraphics[width=0.22\textwidth]{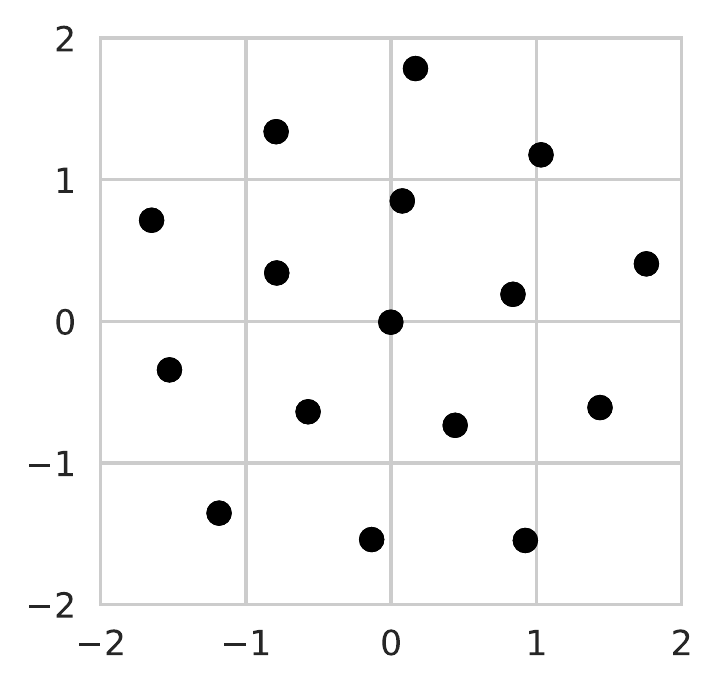} 
\includegraphics[width=0.22\textwidth]{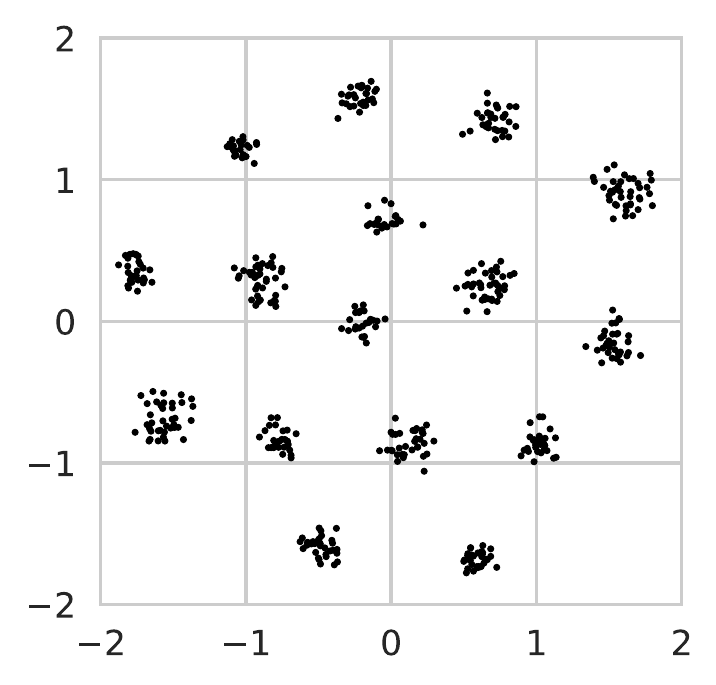} 
\\[-5pt]   
\footnotesize{\hskip4.3mm(c)} \hskip36.8mm  \footnotesize{(d)} 
\end{tabular}
\caption{Constellations produced by autoencoders using parameters $(n,k)$: (a) $(2,2)$ (b) $(2,4)$, (c) $(2,4)$ with average power constraint, (d) $(7, 4)$ 2-dimensional t-SNE embedding of received symbols.}\label{fig:ae_consts}
\end{figure}

Fig.~\ref{fig:ae_consts} shows the learned representations $\xv$ of all messages for different values of $(n,k)$ as complex constellation points, i.e., the $x$- and $y$-axes correspond to the first an second transmitted symbols, respectively. In Fig.~\ref{fig:ae_consts}d for $(7,4)$, we depict the seven-dimensional message representations using a two-dimensional \ac{t-SNE}\cite{maaten2008visualizing} of the noisy observations $\yv$ instead.
Fig.~\ref{fig:ae_consts}a shows the simple $(2,2)$ system which converges rapidly to a classical \ac{QPSK} constellation with some arbitrary rotation.  Similarly, Fig.~\ref{fig:ae_consts}b shows a $(4,2)$ system which leads to a rotated 16-PSK constellation. The impact of the chosen normalization becomes clear from Fig.~\ref{fig:ae_consts}c for the same parameters but with an average power normalization instead of a fixed energy constraint (that forces the symbols to lie on the unit circle). This results in an interesting mixed pentagonal/hexagonal grid arrangement (with indistinguishable \ac{BLER} performance from 16-QAM). 
The constellation has a symbol in the origin surrounded by five equally spaced nearest neighbors, each of which has six almost equally spaced neighbors.
 Fig.~\ref{fig:ae_consts}d for $(7,4)$ shows that the \ac{t-SNE} embedding of $\yv$ leads to a similarly shaped arrangement of clusters.

The examples of this section treat the communication task as a classification problem and the representation of $\sv$ by an $M$-dimensional vector becomes quickly impractical for large $M$. To circumvent this problem, it is possible to use more compact representations of $\sv$ such as a binary vector with $\log_2(M)$ dimensions. In this case, output activation functions such as sigmoid and loss functions such as MSE or binary cross-entropy are more appropriate. 
Nevertheless, scaling such an architecture to very large values of $M$ remains challenging due 
to the size of the training set and model.
We recall that a very important property of the autoencoder is also that it can learn to communicate over \emph{any} channel, even for which no information-theoretically optimal scheme is known.

\begin{table}
\renewcommand{\arraystretch}{1.2} 
\centering
\caption{Layout of the autoencoder used in Figs.~\ref{fig:nn_vs_hamming} and \ref{fig:nn_vs_bpsk}. It has ${(2M+1)(M+n)+2M}$ trainable parameters, resulting in 62, 791, and 135,944 parameters for the (2,2), (7,4), and (8,8) autoencoder, respectively.
}
\label{tab:ae-layout}
\begin{tabular}{l|c}
 Layer    & Output dimensions    \\\hline
 Input & $M$ \\
 Dense + ReLU & $M$ \\
 Dense + linear & $n$ \\
 Normalization & $n$ \\\hline
 Noise & $n$ \\\hline
 Dense + ReLU & $M$ \\
 Dense + softmax & $M$
\end{tabular}
\end{table}

\subsection{Autoencoders for multiple transmitters and receivers}\label{ssec:interference_channel}
\begin{figure}
\centering
\includegraphics[width=0.48\textwidth]{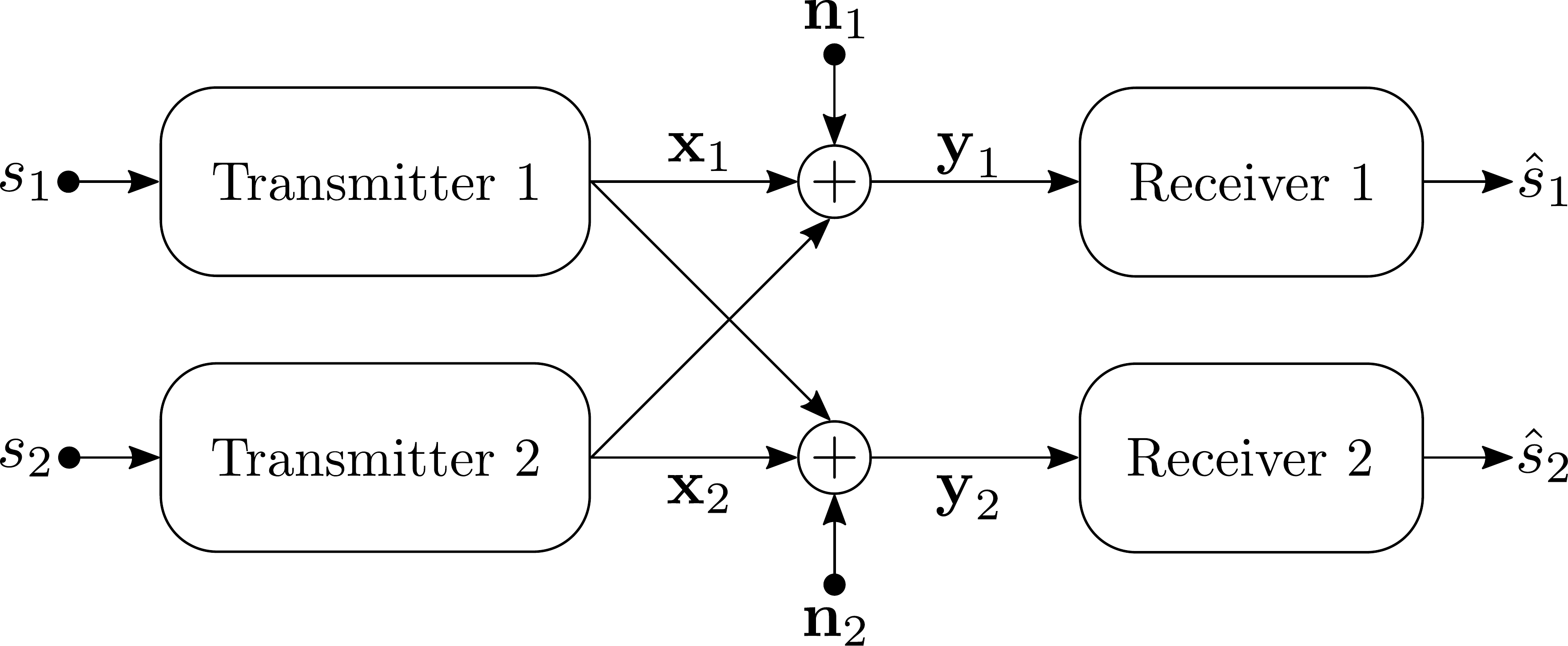}
\caption{The two-user interference channel seen as a combination of two interfering autoencoders that try to reconstruct their respective messages \label{fig:ae_ic}}
\end{figure}

The autoencoder concept from Section~\ref{ssec:ae}, can be readily extended to multiple transmitters and receivers that share a common channel. As an example, we consider here the two-user \ac{AWGN} interference channel as shown in Fig.~\ref{fig:ae_ic}. Transmitter~1 wants to communicate message $s_1\in\mathbb{M}$ to Receiver~1 while Transmitter~2 wants to communicate message $s_2\in\mathbb{M}$ to Receiver~2.\footnote{Extensions to $K$ users with possibly different rates, i.e., $s_k\in\mathbb{M}_k\,\forall k$, as well as to other channel types are straightforward.} Both transmitter-receiver pairs are implemented as \acp{NN} and the only difference with respect to the autoencoder from the last section is that the transmitted messages $\xv_1, \xv_2 \in \mathbb{C}^n$ now interfere at the receivers, resulting in the noisy observations
\begin{align}
 \yv_1 &= \xv_1 + \xv_2 + \nv_1\\
 \yv_2 &= \xv_2 + \xv_1 + \nv_2
\end{align}
where $\nv_1,\nv_1\sim\Cc\Nc(0,\beta\Id_n)$ is Gaussian noise. For simplicity, we have adopted the complex-valued notation, rather than considering real-valued vectors of size $2n$. That is, the notation $(n,k)$ means each of the $2^k$ messages is transmitted over $n$ complex-valued channel uses. Denote by
\begin{equation}
l_1 = -\log\LB\LSB\hat{\sv}_1\RSB_{s_1}\RB, \quad l_2 = -\log\LB\LSB\hat{\sv}_2\RSB_{s_2}\RB
\end{equation}
the individual cross-entropy loss functions of the first and second transmitter-receiver pair, respectively, and by $\tilde{L}_1(\thetav_t)$, $\tilde{L}_2(\thetav_t)$ the associated losses for mini-batch $t$ (cf.~\eqref{eq:def_loss}). In such a context, it is less clear how one should train two coupled autoencoders with conflicting goals. One approach consists of minimizing a weighted sum of both losses, i.e., $\tilde{L} = \alpha \tilde{L}_1 + (1-\alpha) \tilde{L}_2$ for some $\alpha\in[0,1]$. If one would minimize $\tilde{L}_1$ alone (i.e., $\alpha=1$), Transmitter~2 would learn to transmit a constant signal independent of $s_2$ that Receiver~1 could simply subtract from $\yv_1$. The opposite is true for $\alpha=0$. However, giving equal weight to both losses (i.e., $\alpha=0.5$) does not necessarily result in equal performance. We have observed in our experiments that it generally leads to highly unfair and suboptimal solutions. For this reason, we have adopted dynamic weights $\alpha_t$ for each mini-batch $t$:
\begin{align}
\alpha_{t+1} = \frac{\tilde{L}_1(\thetav_t)}{\tilde{L}_1(\thetav_t) + \tilde{L}_2(\thetav_t)},\quad t>0
\end{align}
where $\alpha_0=0.5$. Thus, the smaller $\tilde{L}_1(\thetav_t)$ is compared to $\tilde{L}_2(\thetav_t)$, the smaller is its weight $\alpha_{t+1}$ for the next mini-batch. There are many other possibilities to train such a system and we do not claim any optimality of our approach. However, it has led in our experiments to the desired result of identical \acp{BLER} for both transmitter-receiver pairs.

\begin{figure}
 \centering
 \begin{tikzpicture}
  \begin{semilogyaxis}[
    xmin=0,
    xmax=14,
    ymin=0.00001,
    ymax=1,
    grid=major,
    xlabel={$E_b/N_0$\,[dB]},
    ylabel={Block error rate},
    legend columns=3,
    legend cell align=left,
    legend style={
    		at={(1,1.04)},
    		anchor=south east
    		},
    ]
   \addplot[black, thick] 
    table [x=xa, y=ya, col sep=comma]{ic_vs_timesharing_n1_k1.csv};

 \addplot[black, thick, densely dotted] 
    table [x=xa, y=ya, col sep=comma]{ic_vs_timesharing_n2_k2.csv};

 \addplot[black, thick, densely dashdotted] 
    table [x=xa, y=ya, col sep=comma]{ic_vs_timesharing_n4_k4.csv};
    
 \addplot[red, thick, densely dashdotted, mark=square, mark size=1.5, mark options=solid] 
    table [x=xb, y=yb, col sep=comma]{ic_vs_timesharing_n4_k4.csv};  
    
 \addplot[black, thick, dashed] 
    table [x=xa, y=ya, col sep=comma]{ic_vs_timesharing_n4_k8.csv};  

 \addplot[red, thick, dashed,  mark=o, mark size=1.5, mark options=solid] 
    table [x=xb, y=yb, col sep=comma]{ic_vs_timesharing_n4_k8.csv};

\legend{{TS/AE $(1,1)$}, {TS/AE $(2,2)$}, {TS $(4,4)$}, {AE $(4,4)$}, {TS $(4,8)$}, {AE $(4,8)$}}

  \end{semilogyaxis}
 \end{tikzpicture}
 \caption{\ac{BLER} versus $E_b/N_0$ for the two-user interference channel achieved by the autoencoder (AE) and $2^{2k/n}$-\ac{QAM} time-sharing (TS) for different parameters $(n,k)$}\label{fig:ic_vs_timesharing}
\end{figure}
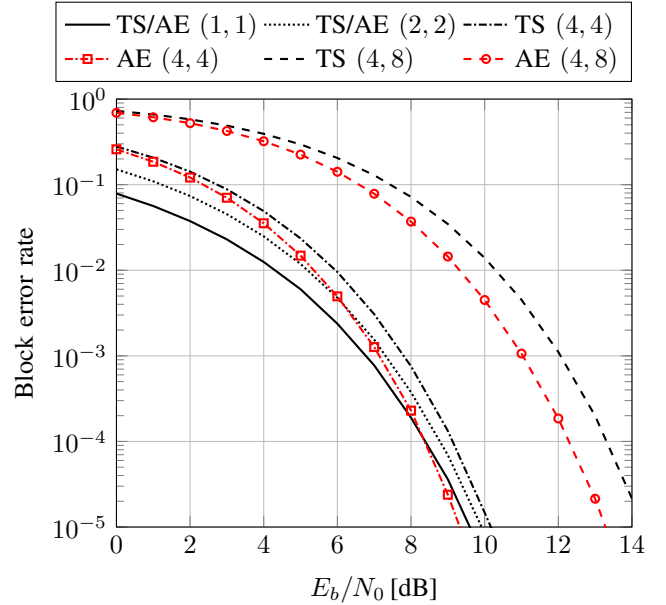

Fig.~\ref{fig:ic_vs_timesharing} shows the \ac{BLER} of one of the autoencoders (denoted by AE) as a function of $E_b/N_0$ for the sets of parameters $(n,k)=\{(1,1), (2,2), (4,4), (4,8)\}$. The \ac{NN}-layout for both autoencoders is that provided in Table~\ref{tab:ae-layout} by letting $n=2n$. We have used an average power constraint to be competitive with higher-order modulation schemes (cf.~Fig.~\ref{fig:ae_consts}c). As a baseline for comparison, we provide the \ac{BLER} of uncoded $2^{2k/n}$-\ac{QAM} that has the same rate when used together with time-sharing (TS) between both transmitters.\footnote{For $(1,1)$, $(2,2)$, and $(4,4)$, each transmitter sends a 4-\ac{QAM} (i.e., \ac{QPSK}) symbol on every other channel use. For $(4,8)$, 16-\ac{QAM} is used instead.} While the autoencoder and time-sharing have identical \acp{BLER} for $(1,1)$ and $(2,2)$, the former achieves substantial gains of around $0.7\,$dB for $(4,4)$ and $1\,$dB for $(4,8)$ at a \ac{BLER} of $10^{-3}$. The reasons for this are similar to those explained in Section~\ref{ssec:ae}. 

\begin{figure}
\centering
\begin{tabular}{l}
  \hskip-3mm\includegraphics[width=0.125\textwidth]{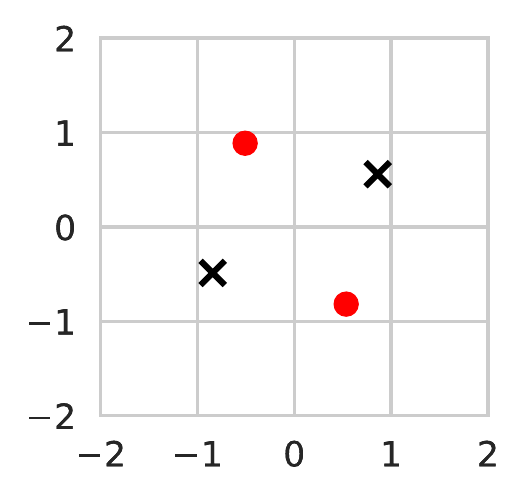}        \hskip7mm\includegraphics[width=0.125\textwidth]{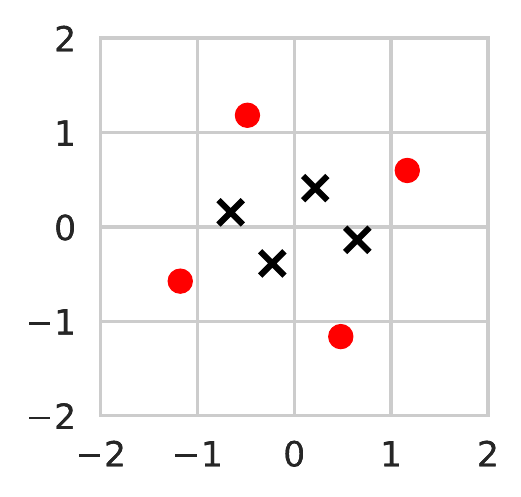} 
   \hskip-2mm\includegraphics[width=0.125\textwidth]{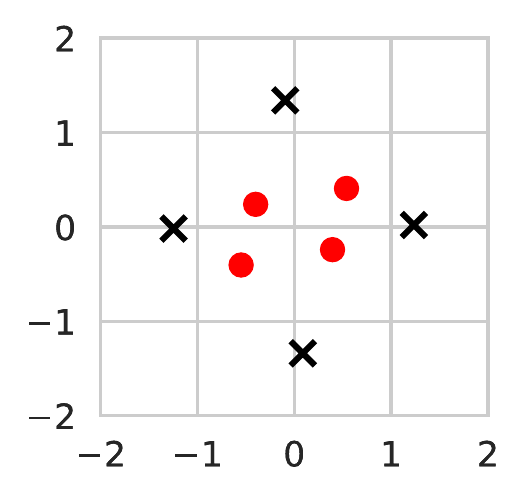} \\[-5pt]   
 \hskip8mm \footnotesize{(a)} \hskip37mm  \footnotesize{(b)}\\[-8pt] 
 \\
\hskip-3mm\includegraphics[width=0.125\textwidth]{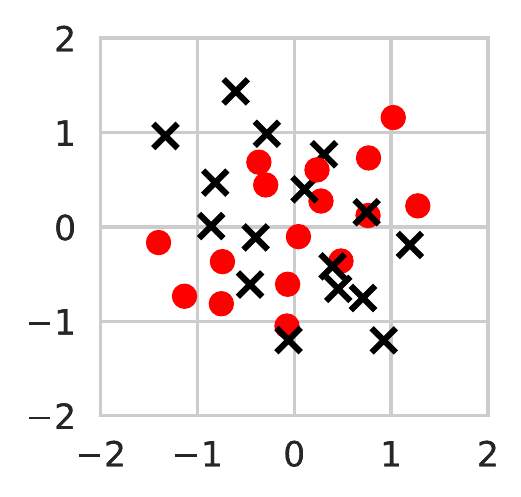}         \hskip-2mm\includegraphics[width=0.125\textwidth]{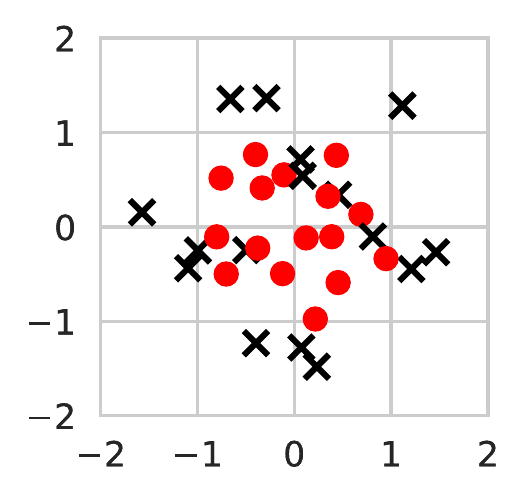} 
    \hskip-2mm\includegraphics[width=0.125\textwidth]{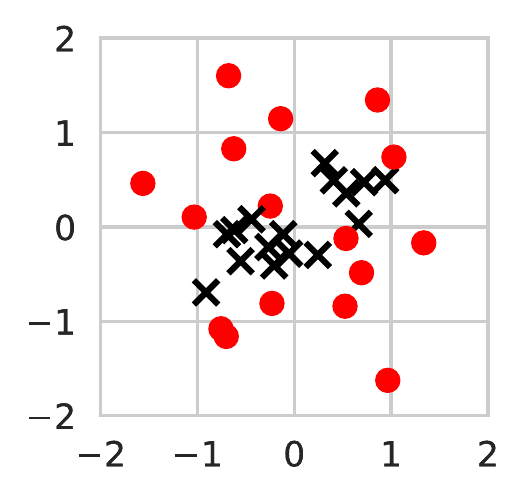}
    \hskip-2mm\includegraphics[width=0.125\textwidth]{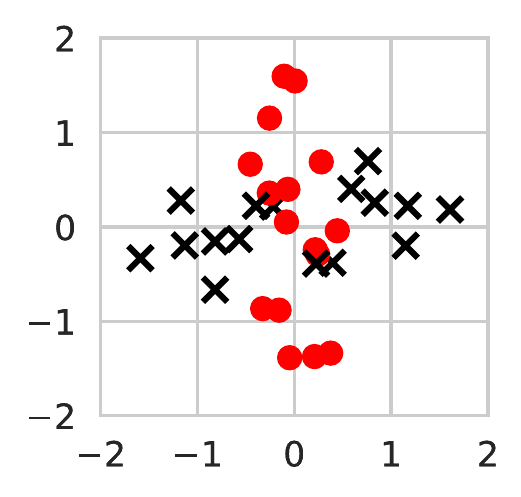}\\[-5pt]
     \hskip40mm \footnotesize{(c)}
      \\
\hskip-3mm\includegraphics[width=0.125\textwidth]{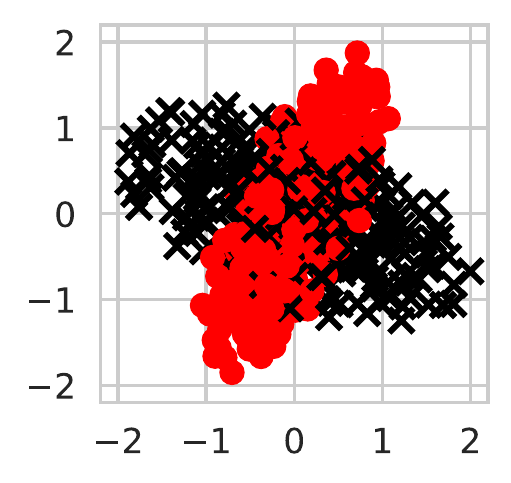}         \hskip-2mm\includegraphics[width=0.125\textwidth]{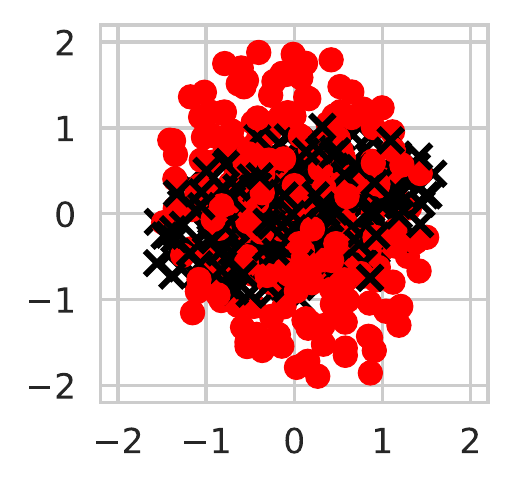} 
    \hskip-2mm\includegraphics[width=0.125\textwidth]{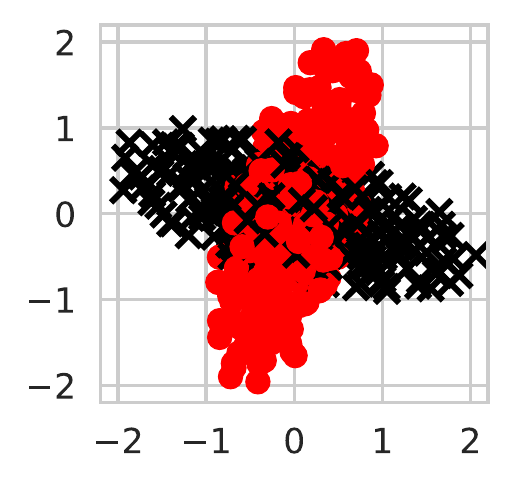}
    \hskip-2mm\includegraphics[width=0.125\textwidth]{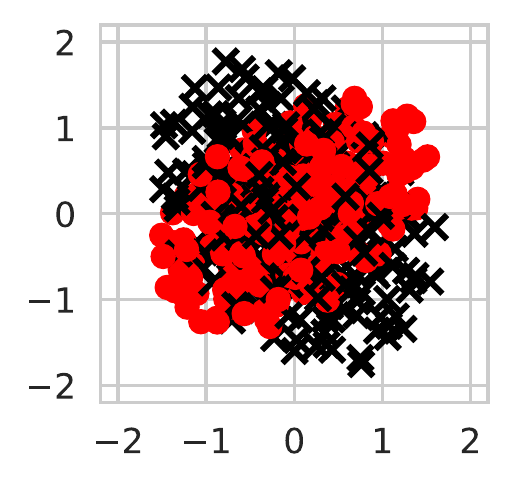}\\[-5pt]
     \hskip40mm \footnotesize{(d)}
\end{tabular}
\caption{Learned constellations for the two-user interference channel with parameters (a) $(1,1)$, (b) $(2,2)$, (c) $(4,4)$, and (d) $(4,8)$. The constellation points of Transmitter~1 and 2 are represented by red dots and black crosses, respectively.}\label{fig:ic_constellations}
\end{figure}

It is interesting to have a look at the learned message representations which are shown in Fig.~\ref{fig:ic_constellations}. For $(1,1)$, the transmitters have learned to use \ac{BPSK}-like constellations in orthogonal directions (with an arbitrary rotation around the origin). This achieves the same performance as \ac{QPSK} with time-sharing. However, for $(2,2)$, the learned constellations are not orthogonal anymore and can be interpreted as some form of super-position coding. For the first symbol, Transmitter~1 uses high power and Transmitter~2 low power. For the second symbol, the roles are changed. For $(4,4)$ and $(4,8)$, the constellations are more difficult to interpret, but we can see that the constellations of both transmitters resemble ellipses with orthogonal major axes and varying focal distances. This effect is more visible for $(4,8)$ than for $(4,4)$ because of the increased number of constellation points. An in-depth study of learned constellations and how they are impacted by the chosen normalization and \ac{NN} weight initializations is out of the scope of this paper but a very interesting topic of future investigations.

We would like to point out that one can easily consider other types of multi-transmitter/receiver communications systems with this approach. These comprise, the general \ac{MAC} and \ac{BC}, as well as systems with jammers and eavesdroppers. As soon as some of the transmitters and receivers are non-cooperative, adversarial training strategies could be adopted (see \cite{goodfellow2014generative,abadi2016learning}).

\subsection{Radio transformer networks for augmented signal processing algorithms}\label{ssec:rtn}
\begin{figure*}
\centering
\includegraphics[width=0.65\textwidth]{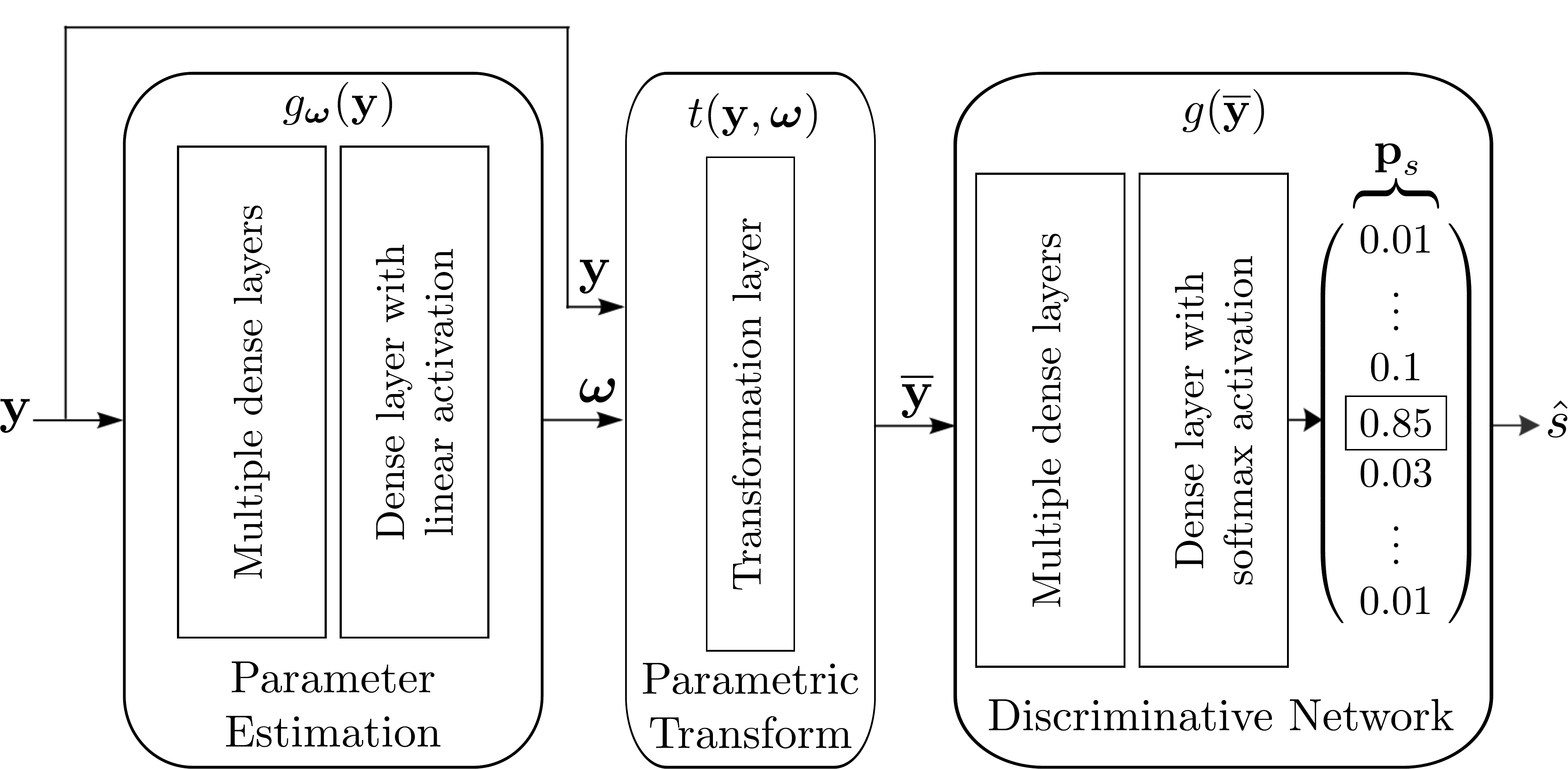}
\caption{A radio receiver represented as an RTN. The input $\yv$ first runs through a parameter estimation network $g_\omega(\yv)$, has a known transform $t(\yv,\omegav)$ applied to generate the canonicalized signal $\overline{\yv}$, and then is classified in the discriminative network $g(\overline{\yv})$ to produce the output $\hat{s}$. \label{fig:rtn}}
\end{figure*}

Many of the physical phenomena undergone in a communications channel and in transceiver hardware can be inverted using compact parametric models/transformations. 
Widely used transformations include re-sampling to estimated symbol/clock timing, mixing with an estimated carrier tone, and convolving with an inverse channel impulse response.
The estimation processes for parameters to seed these transformations (e.g., frequency offset, symbol timing, impulse response) is often very involved and specialized based on signal specific properties and/or information from pilot tones (see, e.g., \cite{meyr1998digital}). 

One way of augmenting \ac{DL} models with expert propagation domain knowledge but not signal specific assumptions is through the use of an \ac{RTN} as shown in Fig.~\ref{fig:rtn}. An RTN consists of three parts: (i) a learned parameter estimator $g_{\boldsymbol\omegav}:\RR^{n} \mapsto \RR^{p}$ which computes a parameter vector $\omegav\in \RR^{p}$ from its input $\yv$, (ii) a parametric transform $t: \RR^{n} \times \RR^{p}\mapsto \RR^{n'}$ that applies a deterministic (and differentiable) function to $\yv$ which is parametrized by $\omegav$ and suited to the propagation phenomena, and (iii) a learned discriminative network $g:\RR^{n'} \mapsto \Mc$ which produces the estimate $\hat{s}$ of the transmitted message (or other label information) from the canonicalized input $\overline{\yv}\in\RR^{n'}$. By allowing the parameter estimator $g_{\boldsymbol\omegav}$ to take the form of an \ac{NN}, we can train the system end-to-end to optimize for a given loss function. Importantly, the training process of such an \ac{RTN} does not seek to directly improve the parameter estimation itself but rather optimizes the way the parameters are estimated to obtain the best end-to-end performance (e.g., \ac{BLER}). While the example above describes an \ac{RTN} for receiver-side processing, it can similarly be used wherever parametric transformations seeded by estimated parameters are needed. \acp{RTN} are a form of learned feed-forward attention inspired by Spatial Transformer Networks (STNs) \cite{jaderberg2015spatial} which have worked well for computer vision problems. 

The basic functioning of an \ac{RTN} is best understood from a simple example, such as the problem of phase offset estimation and compensation. Let $\yv_c=e^{j\varphi}\tilde{\yv}_c\in\CC^{n}$ be a vector of IQ samples that have undergone a phase rotation by the phase offset $\varphi$, and let $\yv=[\Re\{\yv\}\tp, \Im\{\yv\}\tp]\tp\in\RR^{2n}$. The goal of $g_{\boldsymbol\omegav}$ is to estimate a scalar $\hat{\varphi} = \omegav= g_{\boldsymbol\omegav}(\yv)$ that is close to the phase offset $\varphi$, which is then used by the parametric transform $t$ to compute $\bar{\yv}_c = e^{-j\hat{\varphi}}\yv_c$. The canonicalized signal $\bar{\yv}=[\Re\{\bar{\yv}_c\}\tp, \Im\{\bar{\yv}_c\}\tp]\tp$ is thus given by 
\begin{align}
\bar{\yv} &= t(\hat{\varphi}, \yv)= \begin{bmatrix}
\cos(\hat{\varphi})\Re\{\bar{\yv}_c\} + \sin(\hat{\varphi})\Im\{\bar{\yv}_c\}\\
\cos(\hat{\varphi})\Im\{\bar{\yv}_c\} - \sin(\hat{\varphi})\Re\{\bar{\yv}_c\}
\end{bmatrix}
\end{align}
and then fed into the discriminative network $g$ for further processing, such as classification.  

A compelling example demonstrating the advantages of \acp{RTN} is shown in Fig.~\ref{fig:rayleigh} which compares the \ac{BLER} of an autoencoder (8,4)\footnote{We assume complex-valued channel uses, so that transmitter and receiver have $2n$ real-valued inputs and outputs.} with and without \ac{RTN} over a multipath fading channel with $L=3$ channel taps. That is, the received signal $\yv=[\Re\{\yv_c\}\tp,\Im\{\yv_c\}\tp]\tp\in\RR^{2n}$ is given as 
\begin{equation}\label{eq:rayleigh_channel}
y_{c,i}=\sum_{\ell=1}^{L}h_{c,\ell}x_{c,i-\ell+1}+n_{c,i}
\end{equation}
where ${\hv_c\sim\Cc\Nc(0,L^{-1}\Id_L)}$ are i.i.d.\@ Rayleigh fading channel taps, ${\nv_c\sim\Cc\Nc(0,(RE_b/N_0)^{-1}\Id_n)}$ is receiver noise, and $\xv_c\in\CC^n$ is the transmitted signal, where we assume in \eqref{eq:rayleigh_channel} $x_{c,i}=0$ for $i\le 0$. 
Here, the goal of the parameter estimator is to predict a complex-valued vector $\omegav_c$ (represented by $2L$ real values) that is used in the transformation layer to compute the complex convolution of $\yv_c$ with $\omegav_c$. Thus, the \ac{RTN} tries to equalize the channel output through inverse filtering in order to simplify the task of the discriminative network. We have implemented the estimator as an \ac{NN} with two dense layers with $\tanh$ activations followed by a dense output layer with linear activations.  

\begin{figure}
 \centering
 \begin{tikzpicture}
  \begin{axis}[
    xmin=0,
    xmax=100,
    ymin=0,
    ymax=1,
    grid=major,
    xlabel={Training epoch},
    ylabel={Categorical cross-entropy loss},
    legend pos=north east,
    legend cell align={left},
    ]
   \addplot[blue, thick] 
    table [x=xa, y=ya, col sep=comma]{autoencoder_rtn_L3_8_4_training_curve.csv};
    \addplot[red, very thick, dashed] 
    table [x=xb, y=yb, col sep=comma]{autoencoder_rtn_L3_8_4_training_curve.csv};
   \legend{{Autoencoder},{Autoencoder + RTN}}
  \end{axis}
 \end{tikzpicture}
 \caption{Autoencoder training loss with and without RTN}\label{fig:rtn_cae}
\end{figure}
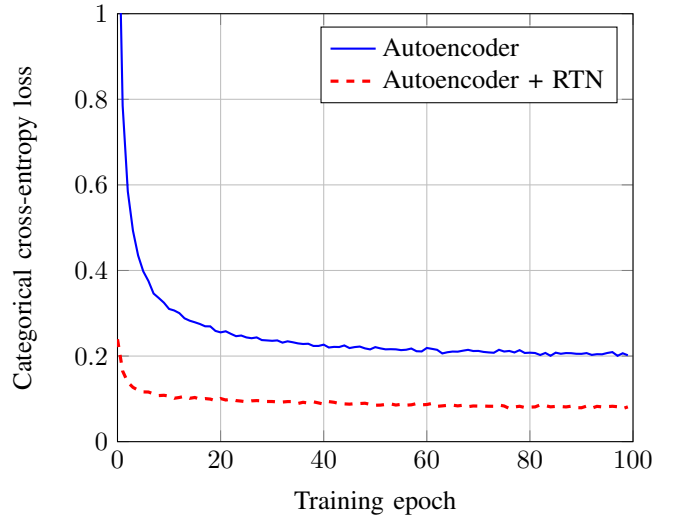

 While the plain autoencoder struggles to meet the performance of differential BPSK (DBPSK) with maximum likelihood sequence estimation (MLE) and a Hamming (7,4) code, the autoencoder with \ac{RTN} outperforms it. Another advantage of \acp{RTN} is faster training convergence which can be seen from Fig.~\ref{fig:rtn_cae} that compares the validation loss of the autoencoder with and without \ac{RTN} as a function of the training epochs. We have observed in our experiments that the autoencoder with \ac{RTN} consistently outperforms the plain autoencoder, independently of the chosen hyper-parameters. However, the performance differences diminish when the encoder and decoder networks are made wider and trained for more iterations. Although there is theoretically nothing an \ac{RTN}-augmented \ac{NN} can do that a plain \ac{NN} cannot, the \ac{RTN} helps by incorporating domain knowledge to simplify the target manifold, similar to the role of convolutional layers in imparting translation invariance where appropriate. This leads to a simpler search space and improved generalization. 

The autoencoder and \ac{RTN} as presented above can be easily extended to operate directly on IQ samples rather than symbols to effectively deal with problems such as pulse shaping, timing-, frequency- and phase-offset compensation. This is an exciting and promising area of research that we leave to future investigations. Interesting applications of this approach could also arise in optical communications dealing with highly non-linear channel impairments that are notoriously difficult to model and compensate for \cite{estaran2016artificial}.
 
\begin{figure}
 \centering
 \begin{tikzpicture}
  \begin{semilogyaxis}[
    xmin=0,
    xmax=20,
    ymin=0.0001,
    ymax=1,
    grid=major,
    xlabel={$E_b/N_0$\,[dB]},
    ylabel={Block error rate},
    legend pos=south west,
    legend cell align={left},
    ]
   \addplot[black, mark=*, thick] 
    table [x=xa, y=ya, col sep=comma]{autoencoder_rtn_L3_8_4.csv};
   \addplot[red, mark=o, mark options={solid}, thick, dashed] 
    table [x=xb, y=yb, col sep=comma]{autoencoder_rtn_L3_8_4.csv};   
   \addplot[blue, mark=square, thick] 
    table [x=xc, y=yc, col sep=comma]{autoencoder_rtn_L3_8_4.csv};      
   \legend{{Autoencoder (8,4)}, {DBPSK(8,7) MLE + Hamming(7,4)}, {Autoencoder (8,4) + RTN}, {Coherent BPSK (8,8)}}
     \end{semilogyaxis}
 \end{tikzpicture}
 \caption{BLER versus $E_b/N_0$ for various communication schemes over a channel with $L=3$ Rayleigh fading taps}\label{fig:rayleigh}
\end{figure}
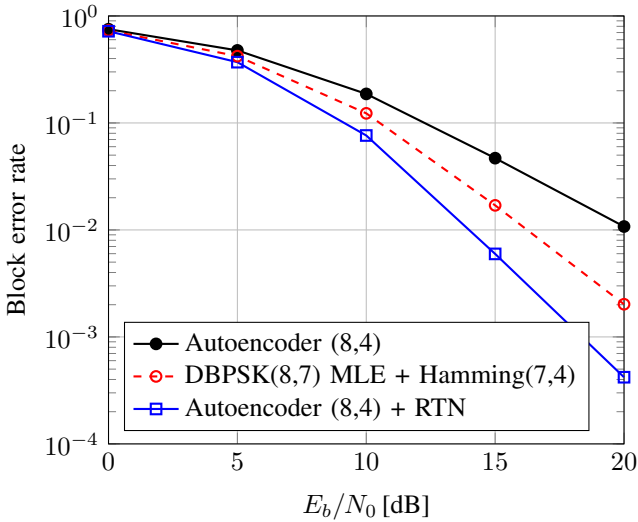

\subsection{CNNs for  classification tasks}\label{ssec:classification}

Many signal processing functions within the physical layer can be learned as either regression or classification tasks. Here we look at the well-known problem of modulation classification of single carrier modulation schemes based on sampled radio frequency time-series data, i.e., IQ samples. This task has been accomplished for years through the approach of expert feature engineering and either analytic decision trees (single trees are widely used in practice) or trained discrimination methods operating on a compact feature space, such as support vector machines, random forests, or small feedforward \acp{NN} \cite{nandi1998algorithms}. Some recent methods take a step beyond this using pattern recognition on expert feature maps, such as the spectral coherence function or $\alpha$-profile, combined with \ac{NN}-based classification \cite{fehske2005new}. However, approaches to this point have not sought to use feature learning on raw time-series data in the radio domain. This is however now the norm in computer vision which motivates our approach here.

As is widely done for image classification, we leverage a series of narrowing convolutional layers followed by dense/fully-connected layers and terminated with a dense softmax layer for our classifier (similar to a VGG architecture \cite{simonyan2014very}). The layout is provided in Table~\ref{tab:cnn-layout} and we refer to the source code \cite{radioML} for further implementation details. The dataset\footnote{RML2016.10b---\url{https://radioml.com/datasets/radioml-2016-10-dataset/}} for this benchmark consists of $1.2$M sequences of 128 complex-valued basedband IQ samples corresponding to ten different digital and analog single-carrier modulation schemes (AM, FM, PSK, QAM, etc.) that have gone through a wireless channel with harsh realistic effects including multipath fading, sample rate and center frequency offset \cite{o2016convolutional}. The samples are taken at 20 different \acp{SNR} within the range from $-20\,$dB to $18\,$dB.

\begin{figure}
 \centering
 \begin{tikzpicture}
  \begin{axis}[
    xmin=-20,
    xmax=18,
    ymin=0,
    ymax=1,
    grid=major,
    xlabel={SNR},
    ylabel={Correct classification probability},
    legend pos=south east,
    legend cell align={left},
    ]   
   \addplot[blue, thick, mark=o] 
    table [x=xa, y=ya, col sep=comma]{cnn_raw_accuracy.csv};
    \addplot[red, thick, mark=square] 
    table [x=xa, y=ya, col sep=comma]{xgb_expert_accuracy.csv};
    \addplot[green, thick, mark=triangle] 
    table [x=xa, y=ya, col sep=comma]{dtree_expert_accuracy.csv};    
     \addplot[black, thick, densely dashed] plot coordinates {(-20,0.1)(18,0.1)};
     
   \legend{{CNN}, {Boosted Tree},{Single Tree},{Random Guessing}}
  \end{axis}
 \end{tikzpicture}
 \caption{Classifier performance comparison versus \ac{SNR}}\label{fig:classifier}
\end{figure}
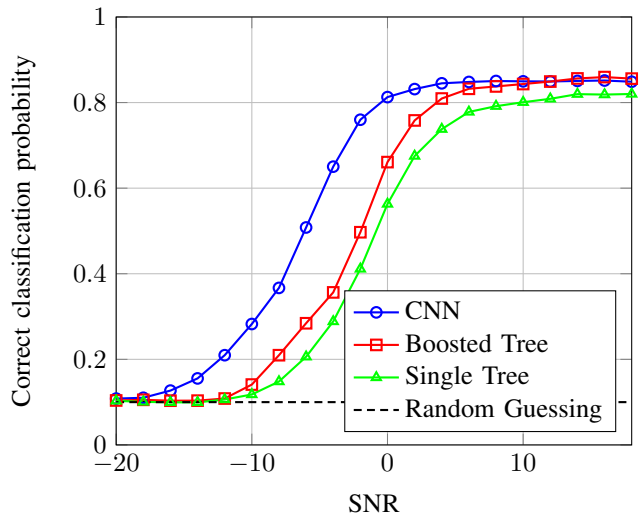

\begin{figure}
\centering
\includegraphics[width=0.422\textwidth]{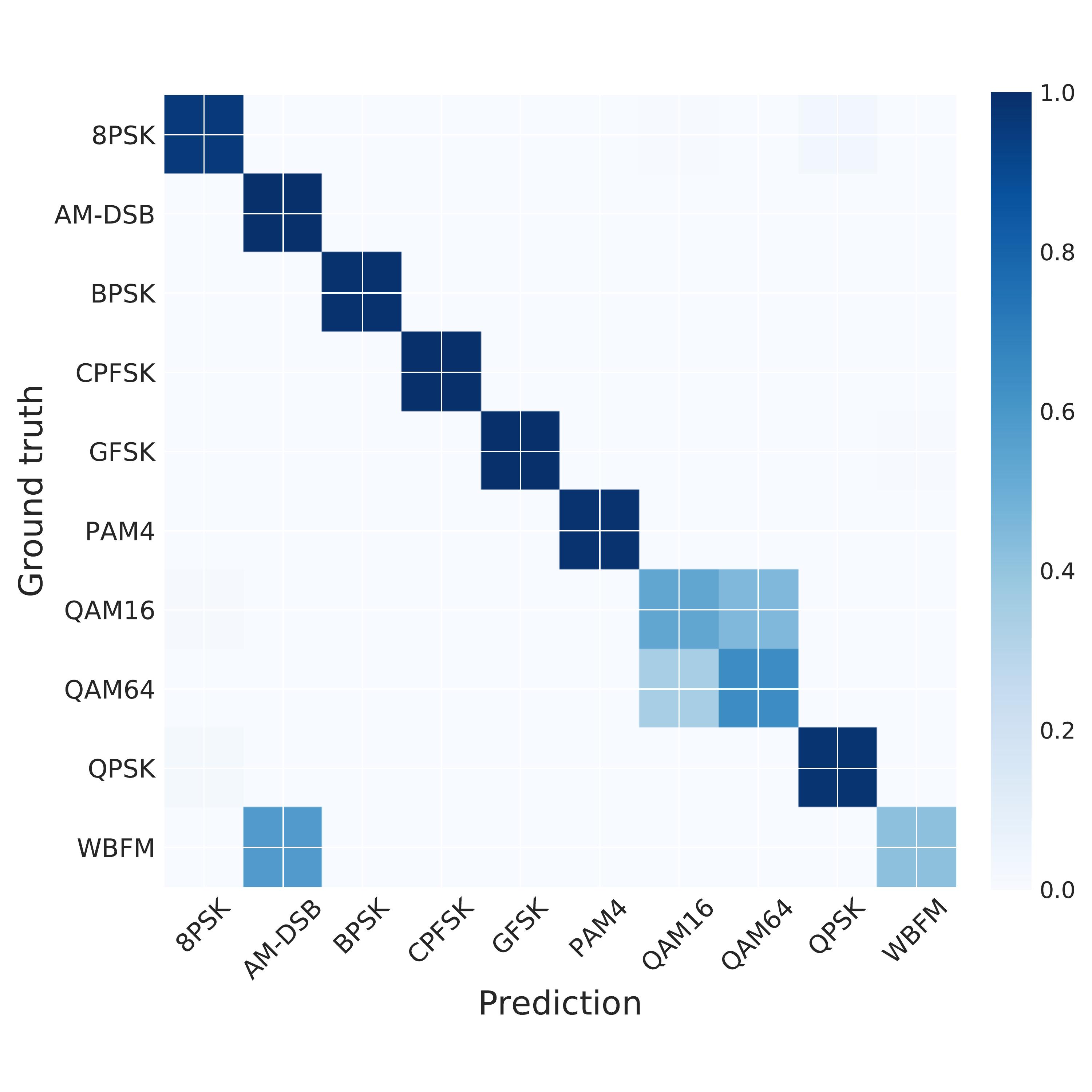}\vskip-4mm
\caption{Confusion matrix of the \ac{CNN}  ($\text{SNR}=10\,$dB)}\label{fig:confusion}
\end{figure}

In Fig.~\ref{fig:classifier}, we compare the classification accuracy of the \ac{CNN} against that of extreme gradient boosting\footnote{At the time of writing of this document, XGB (\url{http://xgboost.readthedocs.io/}) was together with \acp{CNN} the \ac{ML} model that consistently won competions on the data-science platform Kaggle (\url{https://www.kaggle.com/}).} with 1000 estimators, as well as a single scikit-learn tree \cite{scikit-learn}, operating on a mix of 16 analog and cumulant expert features as proposed in \cite{nandi1998algorithms} and \cite{abdelmutalab2016automatic}. The short-time nature of the examples places this task on difficult end of the modulation classification spectrum since we cannot compute expert features with high stability over long periods of time. We can see that the \ac{CNN} outperforms the boosted feature-based classifier by around 4\,dB in the low to medium \ac{SNR} range while the performance at high \ac{SNR} is almost identical. Performance in the single tree case is about $6\,$dB worse than the \ac{CNN} at medium \ac{SNR} and $3.5\,$\% worse at high \ac{SNR}.  Fig.~\ref{fig:confusion} shows the confusion matrix for the \ac{CNN} at $\text{SNR}=10\,\text{dB}$ revealing confusing cases for the \ac{CNN} are between QAM16 and QAM64 and between analog modulations Wideband FM (WBFM) and double-sideband AM (AM-DSB), even at high SNR. The confusion between AM and FM arises during times when the underlying voice signal is idle or does not cary much information. The distinction between QAM16 and QAM64 is very hard with a short-time observation over only a few symbols which share constellation points.
The accuracy of the feature-based classifier saturates at high \ac{SNR} for the same reasons. In \cite{george2016deep}, the authors report on a successful application of a similar \ac{CNN}  for the detection of black hole mergers in astrophysics from noisy time-series data. 

\begin{table}
\renewcommand{\arraystretch}{1.0} 
\centering
\caption{Layout of the \ac{CNN} for modulation classification with 324,330 trainable parameters
}
\label{tab:cnn-layout}
\begin{tabular}{l|l}
 Layer    & Output dimensions    \\\hline
 Input & $2\times 128$ \\
 Convolution (128 filters, size $2\times 8$) + ReLU  & $128 \times 121$ \\
 Max Pooling (size 2, strides 2) & $128 \times 60$\\
 Convolution (64 filters, size $1\times 16$) + ReLU & $64\times 45$ \\
  Max Pooling (size 2, strides 2) & $64 \times 22$\\
 Flatten & $1408$ \\
 Dense + ReLU & $128$ \\
  Dense + ReLU & $64$ \\
 Dense + ReLU & $32$ \\
 Dense + softmax & $10$ 
\end{tabular}
\end{table}

\section{Discussion and open research challenges}
\label{sec:discussion}

\subsection{Data sets and challenges}
In order to compare the performance of ML models and algorithms, it is crucial to have common benchmarks and open datasets. While this is the rule in the computer vision, voice recognition, and natural language processing domains (e.g., MNIST\footnote{\url{http://yann.lecun.com/exdb/mnist/}} or ImageNet\footnote{\url{http://www.image-net.org/}}), nothing comparable exists for communications. 
This domain is somewhat different because it deals with inherently man-made signals that can be accurately generated synthetically, allowing the possibility of standardizing data generation routines rather than just data in some cases.
It would be also desirable to establish a set of common problems and the corresponding datasets (or data-generating software) on which researchers can benchmark and compare their algorithms. One such example task is modulation classification in Section~\ref{ssec:classification}; others could include mapping of impaired received IQ samples or symbols to codewords or bits. Even ``autoencoder competitions'' could be held for a standardized set of benchmark impairments,  taking the form of canonical ``impairment layers'' that would need to be made available for some of the major \ac{DL} libraries (see Section~\ref{sec:ml-lib}).

\subsection{Data representation, loss functions, and training SNR}
\label{ssec:data-loss-snr}
As \ac{DL} for communications is a new field, little is known about optimal data representations, loss-functions, and training strategies. For example, binary signals can be represented as binary or one-hot vectors, modulated (complex) symbols, or integers, and the optimal representation might depend among other factors of the \ac{NN} architecture, learning objective, and loss function. In decoding problems, for instance, one would have the choice between plain channel observations or (clipped) log-likelihood ratios. In general, it seems that there is a representation which is most suited to solve a particular task via an \ac{NN}. Similarly, it is not obvious at which \ac{SNR}(s) \ac{DL} processing blocks should be trained. It is clearly desirable that a learned system operates at any \ac{SNR}, regardless at which \ac{SNR} or \ac{SNR}-range it was trained. However, we have observed that this is generally not the case. Training at low \ac{SNR} for instance does not allow for the discovery of structure important in higher \ac{SNR} scenarios. Training across wide ranges of \ac{SNR} can also severely effect training time.
The authors of \cite{george2016deep} have observed that starting off the training at high \ac{SNR} and then gradually lowering it with each epoch led to significant performance improvements for their application. A related question is the optimal choice of loss function. In Sections~\ref{ssec:ae} to \ref{ssec:rtn}, we have treated communications as a classification problem for which the categorical cross-entropy is a common choice. However, for alternate output data representations, the choice is less obvious. Applying an inappropriate loss function can lead to poor results.

Choosing the right \ac{NN} architecture and training parameters for \ac{SGD} (such as mini-batch size and learning rate) are also important practical questions for which no satisfying hard rules exist. Some guidelines can be found in \cite[Ch.~11]{Goodfellow-et-al-2016-Book}, but methods for how to select such \emph{hyper-parameters} are currently an active area of research and investigation in the \ac{DL} world. Examples include architecture search guided by hyper-gradients and differential hyper-parameters \cite{maclaurin2015gradient} as well as genetic algorithm or particle swarm style optimization \cite{bergstra2012random}.  

\subsection{Complex-valued neural networks}
Owing to the widely used complex baseband representation, we typically deal with complex numbers in communications. Most related signal processing algorithms rely on phase rotations, complex conjugates, absolute values, etc. For this reason, it would be desirable to have \acp{NN} operate on complex rather than real numbers \cite{hirose2006complex}. However, none of the previously described \ac{DL} libraries (see Section~\ref{sec:ml-lib}) currently support this due to several reasons. First, it is possible to represent all mathematical operations in the complex domain with a purely real-valued \ac{NN} of twice the size, i.e., each complex number is simply represented by two real values. For example, an \ac{NN} with a scalar complex input and output connected through a single complex weight, i.e., $y = wx$, where $y,w,x\in\CC$, can be represented as a real-valued \ac{NN} $\yv = \Wm\xv$, where the vectors $\yv,\xv \in \RR^2$ contain the real and imaginary parts of $y$ and $x$ in each dimension and $\Wm\in\RR^{2\times 2}$ is a weight matrix. Note that the real-valued version of this \ac{NN} has four parameters while the complex-valued version has only two. Second, a complication arises in complex-valued \acp{NN} since traditional loss and activation functions are generally not holomorphic so that their gradients are not defined. A solution to this problem is Wirtinger calculus \cite{amin2011wirtinger}. Although complex-valued \acp{NN} might be easier to train and consume less memory, we currently believe that they do not provide any significant advantage in terms of expressive power. Nevertheless, we keep them as an interesting topic for future research.

\subsection{ML-augmented signal processing}
The biggest challenge of learned end-to-end communications systems is that of scalability to large message sets $\Mc$. Already for $k=100$\,bits, i.e., $M=2^{100}$ possible messages, the training complexity is prohibitive since the autoencoder must see at least every message once. Also naive neural channel decoders (as studied in \cite{gruber2017}) suffer from this ``curse of dimensionality'' since they need to be trained on  all possible codewords. Thus, rather than switching immediately to learned end-to-end communications systems or fully replacing certain algorithms by \acp{NN}, one more gradual approach might be that of \emph{augmenting} only specific sub-tasks with \ac{DL}. 
 A very interesting approach in this context is  \emph{deep unfolding} of existing iterative algorithms outlined in \cite{hershey2014deep}. This approach offers the potential to leverage additional side information from training data to improve an existing signal processing algorithm. It has been recently applied in the context of channel decoding and \ac{MIMO} detection \cite{nachmani2016learning, nachmani2017rnn, samuel2017deep}. For instance in \cite{nachmani2016learning}, it was shown that training with a single codeword is sufficient since the structure of the code is embedded in the \ac{NN} through the Tanner graph. The concept of \acp{RTN} as presented in Section~\ref{ssec:rtn} is another way of incorporating both side information from existing models along with information derived from a rich dataset into a \ac{DL} algorithm to improve performance while reducing model and training complexity. 

\subsection{System identification for end-to-end learning}
In Sections~\ref{ssec:ae} to \ref{ssec:rtn}, we have tacitly assumed that the transfer function of the channel is known so that the backpropagation algorithm can compute its gradient. For example, for a Rayleigh fading channel, the autoencoder needs to know during the training phase the exact realization of the channel coefficients to compute how a slight change in the transmitted signal $\xv$ impacts the received signal $\yv$. While this is easily possible for simulated systems, it poses a major challenge for end-to-end learning over real channels and hardware. In essence, the hardware and channel together form a black-box whose input and output can be observed, but for which no exact analytic expression is known a priori. Constructing a model for a black box from data is called system identification \cite{goodwin1977dynamic}, which is widely used in control theory. Transfer learning \cite{pan2010survey} is one appealing candidate for adapting an end-to-end communications system trained on a statistical model to a real-world implementation which has worked well in other domains (e.g., computer vision). An  important related question is that of how one can learn a general model for a wide range of communication scenarios and tasks that would avoid retraining from scratch for every individual setting.

\subsection{Learning from CSI and beyond}
Accurate \ac{CSI} is a fundamental requirement for multi-user \ac{MIMO} communications. For this reason, current cellular communication systems invest significant resources (energy and time) in the acquisition of \ac{CSI} at the base station and user equipment. This information is generally not used for anything apart from precoding/detection or other tasks directly related to processing of the current data frame. Storing and analyzing large amounts of \ac{CSI} (or other radio data)---possibly enriched with location information---poses significant potential for revealing novel big-data-driven physical-layer understanding algorithms beyond immidiate radio environment needs. New applications beyond the traditional scope of communications, such as tracking and identification of humans (through walls) \cite{adib2015capturing} as well as gesture and emotion recognition \cite{zhao2016emotion}, could be achieved using ML on radio signals.  


\section{Conclusion}
\label{sec:conclusions}
We have discussed several promising new applications of \ac{DL} to the physical layer. Most importantly, we have introduced a new way of thinking about communications as an end-to-end reconstruction optimization task using autoencoders to jointly learn transmitter and receiver implementations as well as signal encodings without any prior knowledge. 
Comparisons with traditional baselines in various scenarios reveal extremely competitive \ac{BLER} performance, although the scalability to long block lengths remains a challenge. Apart from potential performance improvements in terms of reliability or latency, our approach can provide interesting insight about the optimal communication schemes (e.g., constellations) in scenarios where the optimal schemes are unknown (e.g., interference channel). We believe that this is the beginning of a wide range of studies into \ac{DL} and \ac{ML} for communications and are excited at the possibilities this could lend towards future wireless communications systems as the field matures. For now, there are a great number of open problems to solve and practical gains to be had. We have identified important key areas of future investigation and highlighted the need for benchmark problems and data sets that can be used to compare performance of different \ac{ML} models and algorithms.

\IEEEtriggeratref{52}
\bibliographystyle{IEEEtran}
\bibliography{IEEEabrv,bibliography}
\end{document}